\documentclass[usegraphicx,usenatbib]{mn2e}
\usepackage{aas_macros} 
\usepackage{amssymb} 
\usepackage{graphicx}
\usepackage{lineno}
\usepackage{comment}
\usepackage{color}
\usepackage[T1]{fontenc}
\usepackage{aecompl}
\newcommand{\Mpcns}{\ensuremath{\,h^{-1}\rm Mpc}}
\newcommand{\Mpc}{\ensuremath{\,h^{-1}\rm Mpc\:}}
\newcommand{\Mrns}{\ensuremath{M_{\rm r}^e - 5 \log_{10}{h}}}
\newcommand{\Mr}{\ensuremath{M_{\rm r}^e - 5 \log_{10}{h}\:}}

\newcommand{\Ms}{\ensuremath{M^*}}

\newcommand{\phis}{\ensuremath{\phi^*}}

\newcommand{\alp}{\ensuremath{\alpha}}
\newcommand{\dddpns}{\ensuremath{\delta_{\rm 8}}}
\newcommand{\dddp}{\ensuremath{\delta_{\rm 8}\:}}
\newcommand{\onepd}{\ensuremath{1+\delta_{\rm 8}\:}}
\newcommand{\onepdns}{\ensuremath{1+\delta_{\rm 8}}}
\newcommand{\lonepd}{\ensuremath{\log_{10}{(1+\delta_{\rm 8})}}}
\newcommand{\Nsns}{\ensuremath{N_{\rm s}}}
\newcommand{\Ns}{\ensuremath{N_{\rm s}\:}}
\newcommand{\rsns}{\ensuremath{r_{\rm s}}}
\newcommand{\rs}{\ensuremath{r_{\rm s}\:}}
\newcommand{\colns}{\ensuremath{(g-r)_0}}
\newcommand{\col}{\ensuremath{(g-r)_0\:}}
\newcommand{\phiuns}{\ensuremath{h^{3}\rm Mpc^{-3}}}
\newcommand{\phiu}{\ensuremath{h^{3}\rm Mpc^{-3}\:}}
\newcommand{\Msuns}{\ensuremath{-5\log_{10} h}}
\newcommand{\Msu}{\ensuremath{-5\log_{10} h\:}}

\begin{document}

\title[GAMA: Dependence of LF on environment and colour]{Galaxy And Mass Assembly (GAMA): The dependence of the galaxy luminosity function on environment, redshift and colour}
\author[Tamsyn McNaught-Roberts et al.]{Tamsyn~McNaught-Roberts,$^1$
Peder~Norberg,$^1$ 
Carlton~Baugh,$^1$ 
Cedric~Lacey,$^1$ 
\newauthor
J.~Loveday,$^2$ 
J.~Peacock,$^3$ 
I.~Baldry,$^4$ 
J.~Bland-Hawthorn,$^5$ 
S.~Brough,$^6$ 
\newauthor
Simon~P.~Driver,$^{7,8}$ 
A.~S.~G.~Robotham,$^7$
J.~A.~V\'azquez-Mata$^2$ \\
$^1$ ICC, Department of Physics, Durham University, South Road, Durham, DH1 3LE, UK\\
$^2$ Astronomy Centre, University of Sussex, Falmer, Brighton BN1 9QH\\
$^3$ Institute for Astronomy, University of Edinburgh, Royal Observatory, Edinburgh EH9 3HJ\\
$^4$ Astrophysics Research Institute, Liverpool John Moores University, IC2, Liverpool Science Park, 146 Brownlow Hill, Liverpool, L3 5RF\\
$^5$ Sydney Institute for Astronomy, School of Physics A28, University of Sydney, NSW 2006, Australia\\
$^6$ Australian Astronomical Observatory, PO Box 915, North Ryde, NSW 1670, Australia\\
$^7$ ICRAR (International Centre for Radio Astronomy Research), The University of Western Australia, 35 Stirling Highway, Crawley, WA 6009, Australia\\
$^8$ School of Physics and Astronomy, University of St Andrews, North Haugh, St Andrews, Fife, KY16 9SS, UK}

\pagerange{\pageref{firstpage}--\pageref{lastpage}}

\maketitle
\label{firstpage}

\begin{abstract}
 We use 80922 galaxies in the Galaxy And Mass Assembly (GAMA) survey to measure the galaxy luminosity function (LF) in different environments over the redshift range $0.04<z<0.26$. The depth and size of GAMA allows us to define samples split by colour and redshift to measure the dependence of the LF on environment, redshift and colour. We find that the LF varies smoothly with overdensity, consistent with previous results, with little environmental dependent evolution over the last 3 Gyrs. The modified GALFORM model predictions agree remarkably well with our LFs split by environment, particularly in the most overdense environments. The LFs predicted by the model for both blue and red galaxies are consistent with GAMA for the environments and luminosities at which such galaxies dominate. Discrepancies between the model and the data seen in the faint end of the LF suggest too many faint red galaxies are predicted, which is likely to be due to the over-quenching of satellite galaxies. The excess of bright blue galaxies predicted in underdense regions could be due to the implementation of AGN feedback not being sufficiently effective in the lower mass halos.
\end{abstract}

\section{Introduction}
The galaxy luminosity function (LF) is a fundamental tool for probing the distribution of galaxies in the observable Universe. Measuring how the LF varies with environment and other galaxy properties can help us to constrain the environmental processes involved in galaxy formation and evolution.

 Large galaxy redshift surveys have allowed accurate measurements of the LF over a large area and depth (e.g.~\citealt{Lin1996,Norberg2002a,Blanton2003b,Loveday2012}), with samples big enough to split by redshift and galaxy property. These large surveys have allowed the measurement of the LF in voids~\citep{Hoyle2005} and over a large range of environments~\citep{Bromley1998,Hutsi2002,Croton2005,Tempel2011}. Splitting these samples by different galaxy properties also allows an accurate analysis of how galaxies behave in these environments (e.g. ~\citealt{Dressler1980}).

Historical studies of the dependence of the LF on environment have been restricted to the comparison of cluster and field galaxies, due to the small number of galaxies observed. 
It has been well established that the LF in clusters is significantly different from that of field galaxies. For example,~\citet{DePropris2003} found that the LF in clusters in the 2dF Galaxy Redshift Survey (2dFGRS,~\citealt{Colless2003}) differs from the field LF~\citep{Madgwick2002}. The cluster LF has a characteristic magnitude (\Ms) that is 0.3 magnitudes brighter, and a faint-end slope (\alp) that is steeper by 0.1 than the field LF.
To measure the LF over a larger range of environments, and to include galaxies in voids, deep and highly complete galaxy surveys are needed.

~\citet{Croton2005} measured the $\rm b_{\rm J}$-band LF for a range of environments in the 2dFGRS, finding no significant variation of the faint-end slope with environment. However, \Ms~varies smoothly with environment being brighter in denser regions. When further splitting samples by spectral type, faint, late-type galaxies dominate void regions, and clusters contain an excess of bright early-types. This dependence of galaxy properties such as colour on environment has previously been found to be stronger than the morphology-density relation described in~\citet{Dressler1980}~\citep[see][]{Blanton2005}.
A comparable analysis by~\citet{Tempel2011}, using Sloan Digital Sky Survey (SDSS)~\citep{Abazajian2009}, reached a similar conclusion, namely that the faint-end slope depends only weakly on environment. Splitting the SDSS sample by morphological type,~\citet{Tempel2011} concluded the environmental dependence is strong for elliptical galaxies, but the LF of spirals is almost independent of environment. They also found that the brightest galaxies are absent from void regions, which instead are mainly populated by spirals. These dominate the faint end of the LF, whereas the bright end is dominated by ellipticals.

Alternatively, the environmental dependence of the LF can be investigated by considering the properties of groups in which galaxies reside.~\citet{Robotham2006} measured the LF for galaxies in the 2PIGG group catalogue~\citep{Eke2004} for different group luminosities, finding the faint-end slope steepens and \Ms~brightens with increasing group luminosity, but these trends flatten for very rich clusters. This trend is visible for the entire population as well as when split by colour. Following on from this work,~\citet{Robotham2010b} investigate how the LF varies as a function of virial mass and group multiplicity. Both the 2PIGG and the~\citet{Yang2005} (SDSS) group catalogues show similar variations of the galaxy LF with these properties.

The measure of density used determines the underlying environment that can be probed, thus helping to identify the key physical processes that shape galaxy formation. Friends-of-friends algorithms (e.g.~\citealt{Davis1982,Eke2004,Robotham2011}) are a good probe of the scales internal to a dark matter halo, whereas fixed sized apertures are a better measure of the large scale environment, essentially tracing the underlying dark matter distribution~\citep{Muldrew2012}. ~\citet{Brough2013} and ~\citet{Wijesinghe2012} both defined local environment as the 5th nearest neighbour surface density when measuring the dependence of the star formation rate on environment in GAMA. The GAMA Group catalogue is constructed by~\citet{Robotham2011} using a friends-of-friends algorithm, to measure how galaxy properties depend on the underlying matter distribution. This is used by~\citet{Alpaslan2013} to construct a catalogue of filaments, probing the large scale structure of the universe, and by V\'azquez-Mata et al., (in prep) to determine how the LF varies with various group properties.

Galaxy formation models have been used to determine the underlying physical processes that shape the LF~\citep{Benson2003a}, particularly the faint end, and to predict how the LF changes with environment~\citep{Benson2003c, Mo2004}. In particular, the influence of halo mass and the physics of galaxy formation in voids have been investigated in some detail~\citep{Peebles2001,MathisWhite2002,Benson2003b}.~\citet{MathisWhite2002} predict that the faint-end slope of the LF steepens in underdense environments. In contrast, ~\citet{Hoyle2005} measured the LF of galaxies in voids in the SDSS and found that the faint-end slope is much shallower than is predicted by galaxy formation models, suggesting a deficit of dwarf galaxies in these extremely underdense regions.

In this analysis the Galaxy And Mass Assembly (GAMA) survey~\citep{Driver2011} is used to investigate how the galaxy LF varies with environment, cosmic time and colour. GAMA is a highly complete survey down to $m_{\rm r}=19.8$. Our work extends the analysis of~\citeauthor{Croton2005} to higher redshifts and much higher sampling and takes advantage of the more extensive photometry of GAMA to further split the galaxy sample by colour. Another novel feature of our analysis is that we use simulated galaxy data to create lightcone mock galaxy catalogues to test our approach. The availability of mock catalogues also allows us to compare our measurements from GAMA against the predictions from theoretical models on an equal footing. 

The data and mock catalogues used in this analysis are described in \S\ref{subsection:GAMA}, and \S\ref{subsection:mocks}. The methods adopted for measuring local environment, determining splits in colour, and measuring the luminosity function are given in \S\ref{section:env_meas} to \S\ref{section:LF}. Our LFs split by environment, redshift and colour are presented in \S\ref{section:Results} and discussed in \S\ref{section:Disc}. We summarize our findings in \S\ref{section:Conc}.

We adopt a standard $\rm \Lambda$CDM cosmology with $\Omega_{M} = 0.25$, $\Omega_{\rm \Lambda}=0.75$ and $H_0 = 100h\rm kms^{-1}Mpc^{-1}$, the same cosmology as is used when constructing the mock catalogues.

\section{Method}
\label{section:Method}
In this section we describe the data and mock catalogues used, along with the k- and evolution corrections to galaxy magnitudes. This is followed by a discussion of the methods implemented to measure galaxy overdensity, colour and the galaxy luminosity function.

\subsection{GAMA DATA}
\label{subsection:GAMA}
The Galaxy And Mass Assembly (GAMA) survey is a multi-wavelength spectroscopic data set, with input catalogue defined in~\citet{Baldry2010}, tiling strategy explained in ~\citet{Robotham2010a}, GAMA survey output for DR1 and DR2 in ~\citet{Driver2011} and Liske et al. in prep respectively, while the spectroscopic pipeline is described in ~\citet{Hopkins2013}. The GAMA Equatorial regions, G09, G12 and G15, are centered on 9h, 12h and 14.5h in right ascension respectively, each covering 5 x 12 $\rm deg^2$ of sky, totaling $\sim$180 $\rm deg^2$. The data set used is from GAMA-II, defined by SDSS DR7 Petrosian magnitudes, limited to $r_{\rm petro}\leq19.8$, a redshift completeness of $\sim98\%$. We use 80922 galaxies ($z\leq0.26$), with good quality redshifts ($NQ \geq 3$;~\citealt{Driver2011}; Liske et al. in prep).

\begin{figure}
\centering
\includegraphics[width=86mm]{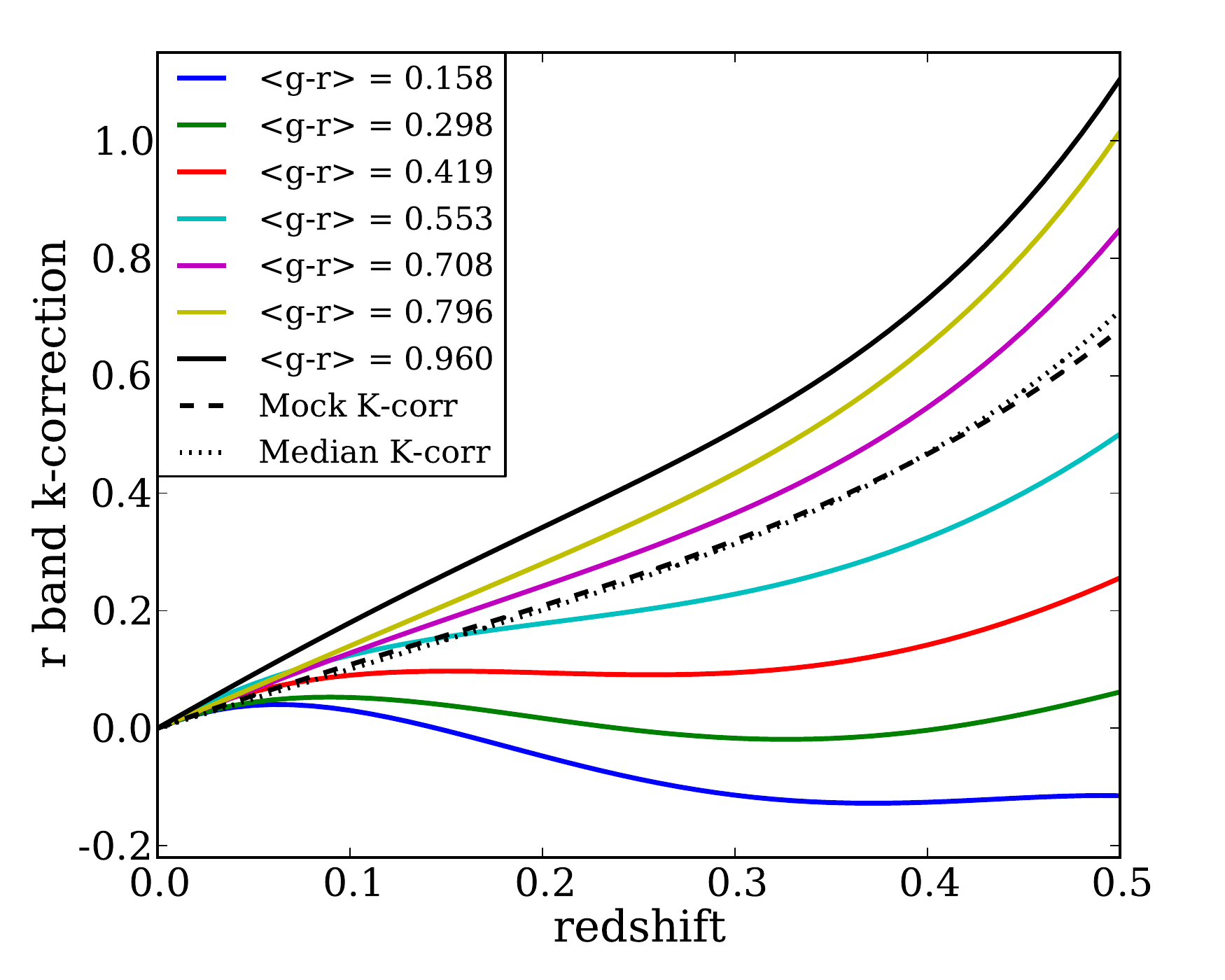}
\caption{\small{Median k-correction tracks to $z_{\rm ref}=0$ for different rest-frame $(g-r)_0$ colours as a function of redshift. The dashed and dotted lines show the k-correction track used for mock galaxies and the median k-correction track of the data. The global k-correction used in the mock catalogues is almost identical to the measured median k-correction for GAMA.}}
\label{fig:kcorcolmed}
\end{figure}

Petrosian magnitudes are k-corrected to account for band shifting when estimating luminosities. This process is described in~\citet{Loveday2012}, and involves fitting an SED to each galaxy using template spectra and SDSS model magnitudes in each of the \emph{ugriz} bands~\citep{Blanton2003a,Blanton2007a}. The redshift dependent k-correction to a reference redshift $z=0$ for each galaxy, $k(z)$, is characterised by a fourth-order polynomial of the form

\begin{equation}
\label{eqn:kcor}
k(z) = \sum\limits_{i=0}^4 a_i (z)^{4-i}.
\end{equation}

To speed up the k-correction calculation, and to account for galaxies with $k(z)$ tracks that differ significantly from the median, thereby over- or underestimating the k-correction of a galaxy at a given redshift, we bin the individual galaxy $k(z)$ into seven bins of uniform width in rest-frame colour \colns. Firstly the $\col$ colour is measured for each galaxy using SDSS \emph{g-} and \emph{r-}band model magnitudes in the observer frame, and individual SED fitted k-corrections for each galaxy. The median $k(z)$ within each \col bin is then calculated ($k_{\rm col}(z)$), and this can be used as an approximate k-correction for all galaxies associated with that bin and at any redshift. The coefficients of the seven colour dependent tracks used in this paper are listed in Table~\ref{tab:kcor} and are shown in Fig.~\ref{fig:kcorcolmed}, together with the median k-correction of the mock catalogues.

%TABLE
\begin{table}
\centering
\begin{tabular}{c c c c c c}
\hline\hline
$(g-r)_{0}$ & $a_{0,col}$ & $a_{1,col}$ & $a_{2,col}$ & $a_{3,col}$ & $a_{4,col}$\\
\hline
$0.158$ & $-31.36$ & $38.63$  & $-14.79$  & $1.427$ & $0.001301$\\
$0.298$ & $-17.77$ & $25.50$  & $-10.79$  & $1.366$ & $0.006235$\\
$0.419$ & $-12.94$ & $21.44$  & $-9.826$  & $1.683$ & $-0.001972$\\
$0.553$ & $-6.299$ & $14.76$  & $-7.473$  & $1.847$ & $-0.006801$\\
$0.708$ & $9.017$  & $-1.390$ & $-0.9145$ & $1.376$ & $-0.004724$\\
$0.796$ & $14.78$  & $-6.592$ & $0.9443$  & $1.357$ & $-0.005131$\\
$0.960$ & $15.09$  & $-5.730$ & $-0.2097$ & $1.859$ & $-0.01250$\\
\hline
\end{tabular}
\caption {\small{median colour, $(g-r)_{0}$, in the seven colour bins and coefficients ($a_{i,col}$ for $i=0,1,2,3,4$) for $k_{\rm col}(z)$ polynomials of the form given in Eqn.~\ref{eqn:kcor}, as shown in Fig.~\ref{fig:kcorcolmed}.}}
\label{tab:kcor}
\end{table}

The luminosity evolution (indicated by $Q_0$) of the sample is taken into account to ensure the sample selection is comparable over a range of redshifts. Luminosity evolution, $E(z)$, is calculated as 

\begin{equation}
\label{eqn:E}
E(z) = -Q_0(z-z_{\rm ref}),
\end{equation}

\noindent where the reference redshift, $z_{\rm ref}$, is the redshift relative to which luminosity evolution is defined ($z_{\rm ref} = 0$). The method implemented to measure $Q_0$ is given in Appendix~\ref{appendix:Q_0}. For all galaxies, we find $Q_{0,\rm all} = 0.97\pm0.15$, and when split into red and blue samples (where colour, \col, is as defined in \S\ref{section:colour}, we find $Q_{0,\rm blue} = 2.12\pm0.22$ and $Q_{0,\rm red} = 0.80\pm0.26$.\footnote{The corresponding $Q_0$ values for mock galaxies are found to be $Q_{0,\rm all} = 0.89\pm0.09$, $Q_{0,\rm blue} = 1.71\pm0.16$ and $Q_{0,\rm red} = 0.63\pm0.07$.}

Petrosian magnitudes ($r_{\rm petro}$) are used to calculate \emph{r}-band absolute magnitudes, as GAMA is selected on $r_{\rm petro}$. The k-corrected and luminosity evolution corrected absolute \emph{r-}band magnitude ($M_{\rm r}^{e}$ at $z=0$) is given by:

\begin{equation}
M_{\rm r}^e - 5 \log_{10} h = r_{\rm petro} - 5 \log_{10} \left(\frac{d_{\rm L} (z)}{\Mpcns}\right) - 25 - k_{col}(z) - E(z)
\end{equation}
with $E(z)$ as given in Eqn.~\ref{eqn:E}, $k_{col}(z)$ depending on galaxy colour and given by Eqn.~\ref{eqn:kcor}, and luminosity distance is given by $d_{\rm L}(z)$. $Q_{0,\rm all}$ is used when defining a volume limited sample (see \S\ref{section:DDP}), while LFs are measured using the specific $Q_{0,\rm red}$ or $Q_{0,\rm blue}$ corresponding to the colour of a galaxy.

\subsection{GAMA Mock Catalogues}
\label{subsection:mocks}
To illustrate how our results can be used to test models of galaxy formation, we perform the same analysis on mock galaxy catalogues. These mock catalogues have the same faint apparent magnitude limit as GAMA, and cover the same area on the sky, allowing a more direct comparison of the properties of the data and the models. The lightcone mock catalogues are constructed from the Millennium dark matter N-body simulation~\citep{Springel2005}, and are populated with galaxies using the~\citet{Bower2006} GALFORM semi-analytic galaxy formation model. For further details of the construction of the mock catalogues, see~\citet{Merson2013}, while a more comprehensive description of the limitations of the GAMA mock catalogues is given in~\citet{Robotham2011}. The \emph{r}-band magnitudes are modified such that the redshift dependent luminosity and selection functions of the mock catalogues match those of GAMA (e.g.~\citealt{Loveday2012}), while the colours and the ranking of galaxies in luminosity remain unchanged. The k-correction track used for mock galaxies is given by Eqn. 8 in~\citet{Robotham2011} and is shown by the dashed black line in Fig.~\ref{fig:kcorcolmed}, very similar to the median track in GAMA (dotted black line). For historical reasons these mock catalogues contain a bright apparent magnitude limit of $m_{\rm r} = 15.0$, restricting the faint luminosity limit of the galaxy luminosity function and the redshift limit over which densities are measured.

The combined mock galaxy catalogue gives better statistics and allows a smoother, more accurate measurement of the galaxy LF. Realistic errors based on the sample variance between the 9 mock catalogues are used to provide error estimates for the mock galaxy LFs.

\subsection{Environment Measure}
\label{section:env_meas}
Environment is defined in terms of galaxy number density smoothed over a localised kernel using a density defining population of galaxies that is introduced in \S\ref{section:DDP}. We explain how the local density of a galaxy is defined in \S\ref{section:dens_meas}.

\subsubsection{Density Defining Population (DDP)}
\label{section:DDP}
A density defining population (DDP) of galaxies is used as a tracer of environment, following~\citet{Croton2005}. This galaxy sample is volume limited given a range of absolute magnitudes ($M_{\rm r}^e$), and the apparent magnitude limits of the survey, that define a limiting redshift range. A galaxy is included as a DDP galaxy if it falls within the absolute magnitude limits of the DDP, and can be seen over the whole redshift range defined by these absolute magnitude limits. 

It is expected that brighter galaxies will reside in denser environments. A brighter DDP sample will therefore cover a larger dynamic range of density in overdense regions, whereas a fainter DDP sample will better sample environments corresponding to underdense regions (i.e. voids). Ideally a DDP sample should cover a large absolute magnitude range, to better sample all environments. However, with a magnitude limited survey, the larger the absolute magnitude range the smaller the range in redshift, and therefore the volume over which overdensities can be measured is reduced. To mitigate sample variance and to enable evolutionary studies, we prefer to use a DDP that covers a reasonably large redshift range, while preserving a high sampling rate.

\begin{figure}
\centering
\includegraphics[width=85mm]{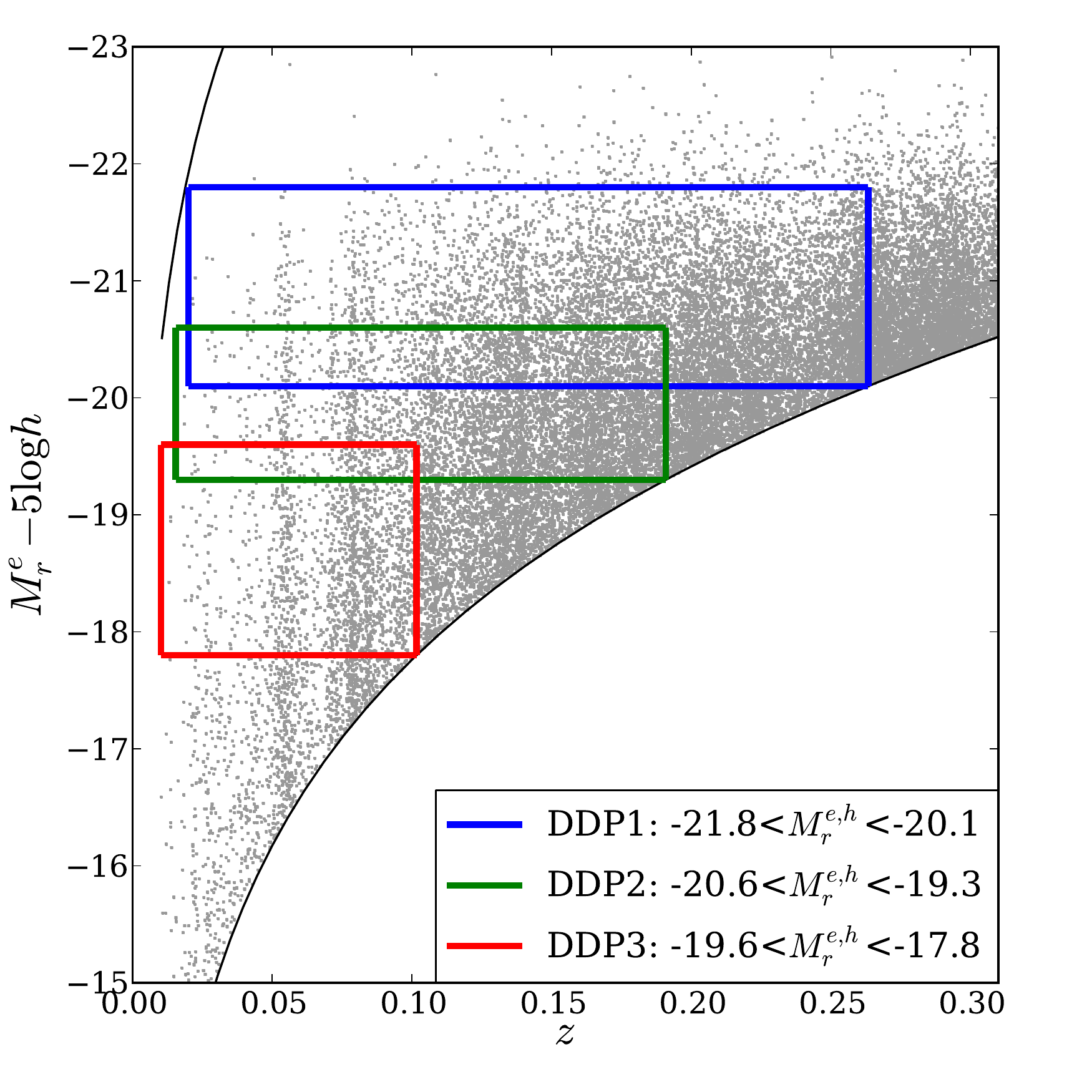}
\caption{\small{Absolute magnitude against redshift for all GAMA data with DDP samples enclosed by different coloured rectangles. Upper and lower black lines show bright and faint apparent magnitude limits of $r=12$ and $r=19.8$ respectively. To define DDP samples a global k-correction is used (see Fig.~\ref{fig:kcorcolmed}). See key for DDP samples, where $M_{\rm r}^{e,h}$ is defined as $M_{\rm r}^e - 5\log_{10}{h}$. DDP1 spans the redshift range $0.04<z<0.26$.}}
\label{fig:M_z}
\end{figure}

Different DDP samples corresponding to different ranges in absolute magnitude and redshift are shown by the coloured rectangles in Fig.~\ref{fig:M_z}, and described in Table~\ref{tab:DDP_samples}. The number of galaxies and subsequently the number density of DDP galaxies is smaller in each of the GAMA DDP samples than in the mock galaxy DDP samples due to redshift incompleteness in GAMA (see \S\ref{section:dens_meas}), which is not modelled in the mock catalogues, and the bright apparent magnitude limit in the mock catalogues, which is fainter in the mock catalogues than in the data, limiting the volume over which densities can be measured. The blue rectangle in Fig.~\ref{fig:M_z}, DDP1, is used to determine the local galaxy environment. It provides a large volume over which environment can be measured and enables evolution with redshift to be investigated. The other DDP samples shown in Fig.~\ref{fig:M_z} and described in Table~\ref{tab:DDP_samples} are used to investigate how robust this measure of environment is, by comparing how the different DDP samples probe the underlying density field.

Once the DDP sample has been defined, all galaxies lying within the redshift limits of the DDP sample can have a local overdensity measured (i.e. including galaxies outside the absolute magnitude range of the DDP). Appendix~\ref{appendix:DDP} compares the overdensity measured using different DDP samples. The measured overdensity does not depend strongly on the DDP sample used, suggesting that this method for measuring environment is fairly insensitive to the precise choice of absolute magnitude range of the density tracers used, once the DDP tracer population is sufficiently dense.

%TABLE
\begin{table*}
\centering
\begin{tabular}{c c c c c r r r r r r}
\hline\hline
DDP&\multicolumn{2}{|c|}{\Mr}&  $z_{\rm min}$  &  $z_{\rm max}$  &  \multicolumn{2}{|c|}{$N_{\rm gal} /10^{3}$}  &  \multicolumn{2}{|c|}{$V_{\rm DDP} /(10^{6} h^{-3} \rm Mpc^{3})$}&\multicolumn{2}{|c|}{$\rho_{\rm DDP} /(10^{-3} h^{3} \rm Mpc^{-3})$}\\
 &faint&bright& & &GAMA&$\langle$Mock$\rangle$&GAMA&$\langle$Mock$\rangle$&GAMA&$\langle$Mock$\rangle$\\
\hline\hline
1& $-20.1$ & $-21.8$ & $0.039$ & $0.263$ & $81.1$ & $84.5 \pm2.3$ & $6.75$ & $6.45 \pm0.02$ &$5.35$ & $6.38 \pm0.18$ \\
2& $-19.3$ & $-20.6$ & $0.015$ & $0.191$ & $47.8$ & $48.3 \pm3.0$ & $2.52$ & $2.42 \pm0.06$ & $8.99$ & $9.47 \pm0.66$ \\
3& $-17.8$ & $-19.6$ & $0.010$ & $0.102$ & $7.88$ & $10.6 \pm2.0$ & $0.32$ & $0.31 \pm0.05$ & $12.7$ & $18.1 \pm6.8$ \\
\hline
\end{tabular}
\caption {\small{Properties of DDP samples. Columns 2-3 list the r-band absolute magnitude range and columns 4-5 list the GAMA redshift ranges. Subsequent columns list the number of galaxies that fall within the DDP redshift limits, the effective co-moving volume of the DDP sample, and the number density of DDP galaxies. For each of these the values for GAMA and the mock catalogues are given, with the latter indicating the mean and scatter from the 9 mock catalogues.}}
\label{tab:DDP_samples}
\end{table*}

\subsubsection{Overdensity}
\label{section:dens_meas}
Once a DDP sample has been defined, the local environment around a galaxy is measured by counting the number of DDP galaxies (\Nsns) that lie within a sphere of a given radius around the galaxy. For this analysis we use a radius of $\rs=8 \Mpc$ (co-moving). Different sphere sizes are discussed in Appendix B of~\citet{Croton2005}, who conclude that smaller spheres ($4 \Mpc$) are a better probe of denser environments. However, sphere sizes that are too small are more likely to be sensitive to redshift-space distortions and shot noise and hence provide less reliable estimates of the density than larger sphere sizes. In agreement with~\citet{Croton2005} we find $8 \Mpc$ radius spheres to be a good probe of both underdense and overdense regions, since larger sphere sizes tend to probe void regions well.

\citet{Muldrew2012} investigate how various measures of environment relate to the underlying dark matter distribution, finding that environment measures using apertures are a better probe of the halo as a whole compared to those using nearest neighbour methods, such that larger density measures more accurately reflect larger halo masses. Larger apertures (e.g. 8\Mpc as used here) correlate well with underlying dark matter environments over large (5\Mpc) scales. However,~\citet{Blanton2007b} compare galaxy properties within the group environment (defined using a friends-of-friends algorithm) to those within a density field over scales ranging from 0.1\Mpc~to 10\Mpc, determining that galaxy properties do not depend on surrounding environment over scales of $>1\Mpc$ any more than the environment within the group.

If a galaxy is close to the edge of the survey, \Ns will be underestimated, as the sphere will sample a volume outside of the survey. This is accounted for by correcting the measured density for the fraction of the sphere volume that falls outside the survey. For an unclustered data set this correction is exact, while for a clustered data set the correction is likely to be less accurate. Spectroscopic completeness is also corrected for in the same way using the GAMA masks. A completeness threshold of 80\% is adopted such that less complete spheres (taking into account redshift and volume completeness) are not included in the analysis (Appendix~\ref{appendix:comp} demonstrates that 77\% of our volume is retained with this cut).

\begin{figure*}
\centering
\includegraphics[width=170mm]{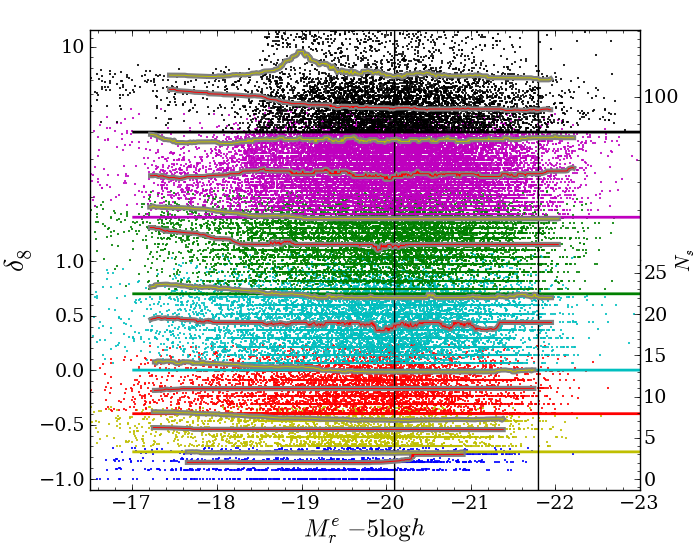}
\caption{\small{Overdensity against absolute magnitude for GAMA data. Black vertical lines show the absolute magnitude limits of the DDP1 sample, solid horizontal lines indicate the lower density limits of our density bins, coloured according to overdensity bin. Each point is coloured according to the overdensity bin it belongs to before completeness corrections are applied. The right side of the y-axis gives the corresponding number of DDP galaxies within an 8\Mpc radius sphere (see \S\ref{section:dens_meas} for discussion). The darker solid lines (red on top of grey) show the running median overdensity (over 1000 galaxies) as a function of absolute magnitude, and the lighter solid lines (yellow on top of grey) show the 90th percentiles. For clarity d2 and d3 are combined here to form the yellow overdensity bin, likewise d7 and d8 are combined to form the magenta overdensity bin. Fainter than $\Mr=-18$, the range over which the running median is calculated is broad ($\sim1$ mag). The y-axis is linear until $\dddp=1$ and logarithmic (base 10) thereafter.}}
\label{fig:M_delta}
\end{figure*}

The local galaxy density, defined within a sphere of radius $r_{\rm s}$, accounting for volume completeness ($C_{\rm v}$) and redshift completeness ($C_{\rm z}$) is given by

\begin{equation}
\rho = \frac{\Ns}{\frac{4}{3} \pi \rsns^3} \frac{1}{C_{\rm v}} \frac{1}{C_{\rm z}},
\end{equation}
for which an overdensity can be calculated for the case $r_{\rm s} = 8 \Mpc$

\begin{equation}
\dddp = \frac{\rho - \overline{\rho}}{\overline{\rho}},
\end{equation}
where $\overline{\rho}$ is the effective mean density of DDP galaxies in the volume. 

Each sample is split into overdensity bins, the basic properties of which are listed in Table~\ref{tab:density_bins} for DDP1. The bins are chosen such that they cover a large range of environments, including extreme underdense and overdense regions where statistics such as the LF may be changing more rapidly. The galaxy LF is measured for all density bins, but for clarity we focus on d1, d4, d6, and d9 from Table~\ref{tab:density_bins}, sampling a variety of environments, from voids (d1) to clusters (d9).

Fig.~\ref{fig:M_delta} shows where galaxies lie in overdensity and absolute magnitude for DDP1, and hence which density bin they fall in (given by solid horizontal lines). Galaxies are coloured according to the density bin they occupy before their local density is corrected for redshift and volume completeness. This shows that there are no significant jumps in density classification: only adjacent bins are affected by the completeness corrections when the threshold of 80\% completeness is imposed. The discrete lines of overdensity (visible especially in the lower density bins) are due to the integer numbers of DDP galaxies within a sphere, corresponding to a specific value of \dddpns. The mean number of DDP galaxies within a $8\Mpc$ radius sphere is 13.2. Galaxies falling between these discrete lines have had their overdensity corrected for incompleteness. 

Since a DDP galaxy will always have at least one galaxy in its overdensity measurement (the DDP galaxy itself is included in $\rm N_{\rm DDP}$), there are no galaxies with $\dddp = -1$ in the magnitude range of the DDP sample (shown by black vertical lines). This effect becomes apparent in the shape of the LF if the lowest density bin considered is chosen to be significantly underdense. To correct for this, the LF estimator in the DDP absolute magnitude range (e.g. between the dashed vertical lines in Fig.~\ref{fig:LF_dens_DDPs_data}) takes into account the effective volume of the DDP sample in each overdensity bin (see \S\ref{section:LF} for details). In the most underdense density bins this volume is much lower for DDP galaxies than for non-DDP galaxies and so not correcting for it would result in an incorrect LF estimate. An alternative approach would be to subtract one from the DDP count when measuring overdensity for a DDP galaxy. However this method implies that the definition of overdensity measured at a position infinitely close to a DDP galaxy is different to that measured at any other position. In order to produce a overdensity measurement which is consistent for all galaxies we use the method described above. This different treatment of DDP galaxies only has significant effect when dealing with small numbers of galaxies in an $8\Mpc$ radius sphere. As Fig.~\ref{fig:M_delta} shows, this is only the case in the lowest density bin, where the correction to the LF as described above is most significant.

 The apparent absence of galaxies at faint magnitudes in the highest overdensity bin plotted in Fig.~\ref{fig:M_delta} is due to this bin being affected by one large cluster in G15 at $z \simeq 0.14$. Given the faint apparent magnitude limit of GAMA and the redshift of the cluster, it is not possible to pick up galaxies fainter than $\Mrns=-18.5$. Most galaxies in this overdensity belong to the largest group recovered in the GAMA group catalogue~\citep{Robotham2011}.

%FIG
\begin{figure*}
\centering
\includegraphics[width=170mm]{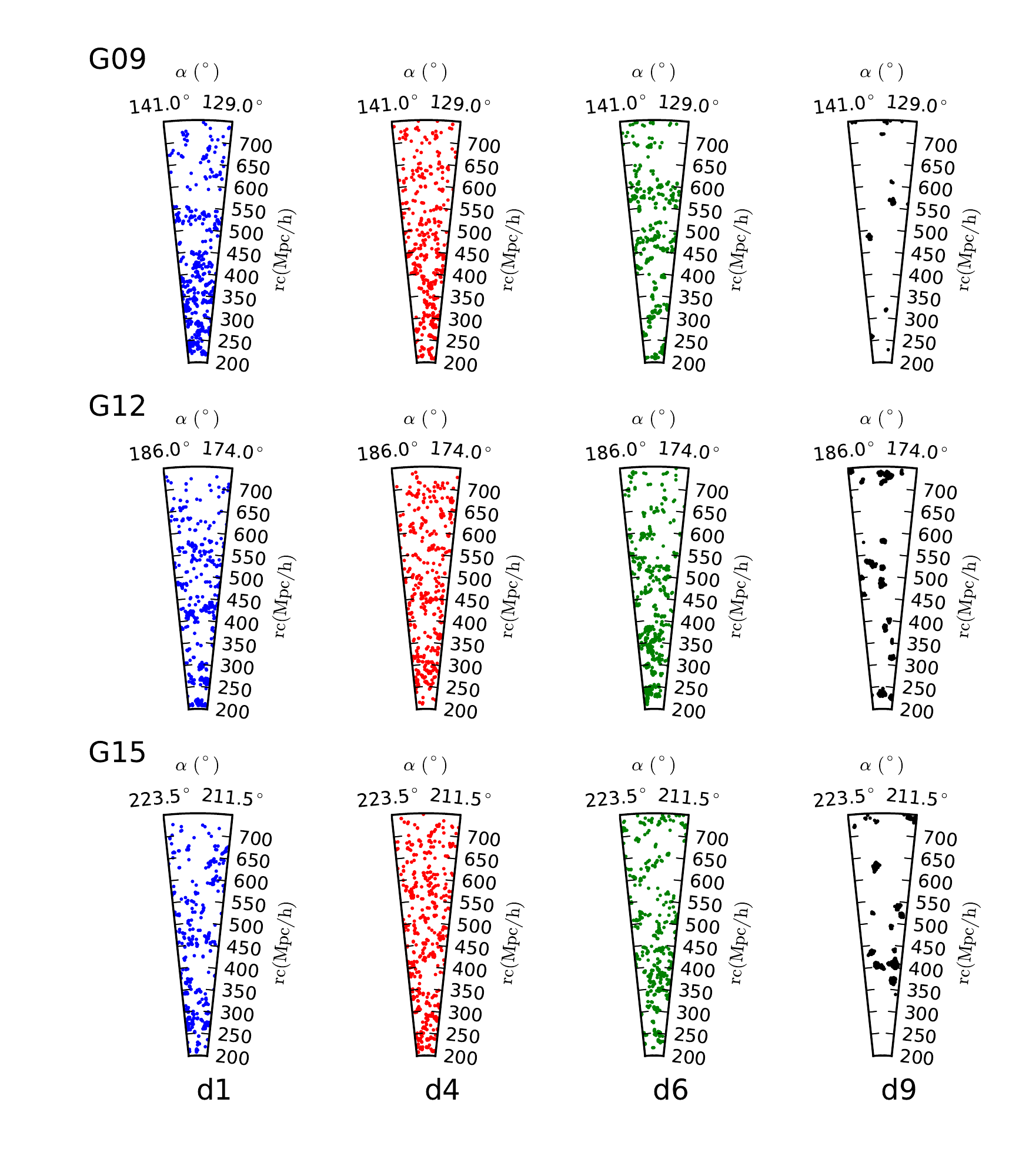}
\caption{\small{The spatial distribution of galaxies for different overdensities (left = most underdense to right = most overdense) in GAMA fields G09, G12, and G15 (top to bottom), over a constant projection thickness of 18.1\Mpcns. Points are coloured according to overdensity bin and are plotted such that a random selection of galaxies totalling the same number in each overdensity bin is shown. Sample variance between the 3 GAMA fields is easily visible, so LFs are estimated using all 3 fields combined.}}
\label{fig:cone_dens}
\end{figure*}

%FIG
\begin{figure*}
\centering
\includegraphics[width=170mm]{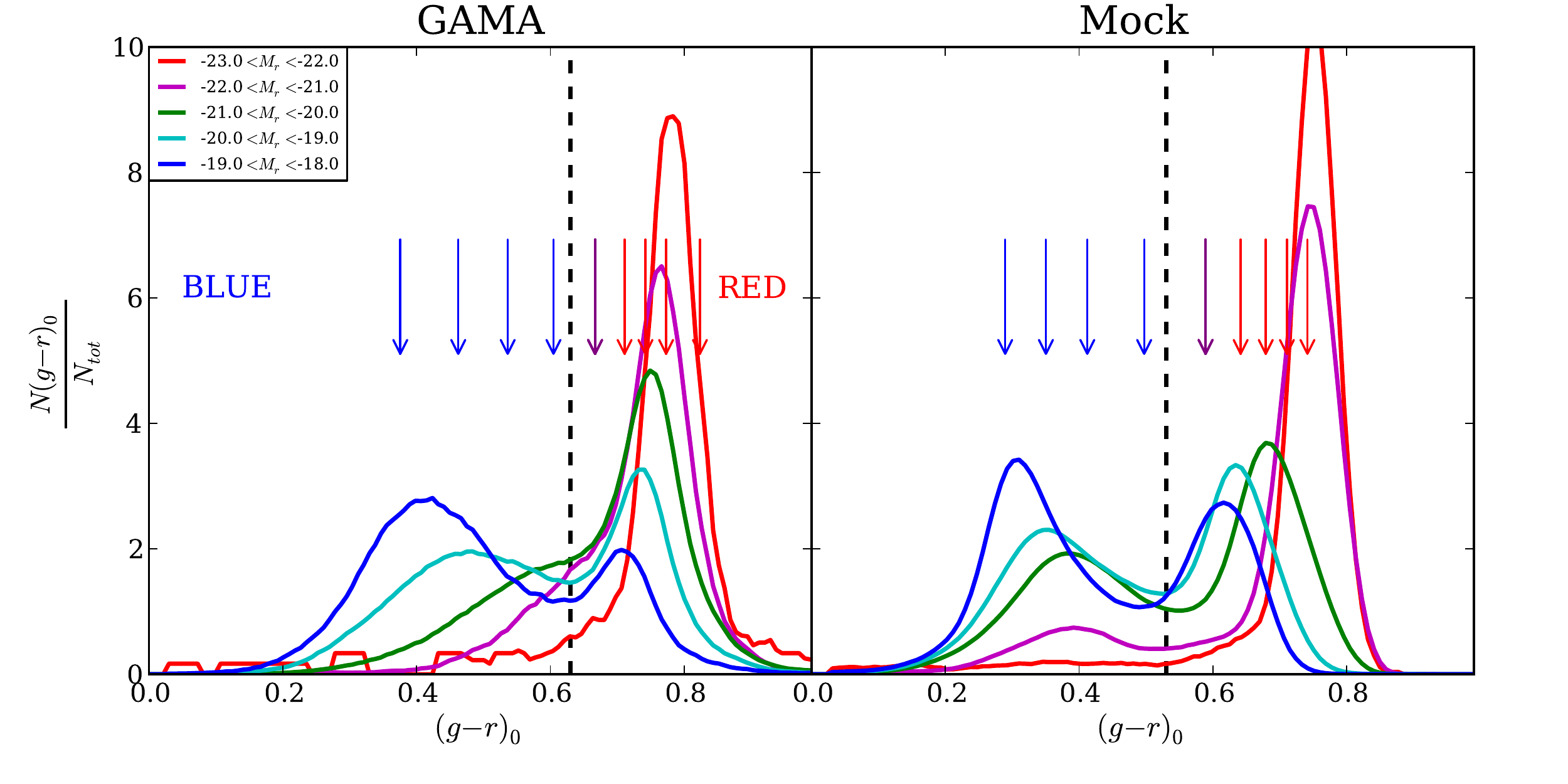}
\caption{\small{Distribution of rest-frame $\colns\:$ colour for 5 different ranges of \emph{r}-band absolute magnitude for GAMA (\emph{left}) and the mock catalogues (\emph{right}). The vertical dashed black lines show the splits in colour used for GAMA and the mock catalogues. The colour split for the mock catalogues is chosen to keep the same fraction of galaxies in each colour sample as for GAMA, whilst ensuring the bimodality in the distribution is still clearly apparent. The arrows correspond to every $10^{\rm th}$ percentile in global \col distribution (see Fig.~\ref{fig:LF_dens_c_data} for results using these splits).}}
\label{fig:g_r_Mg}
\end{figure*}

The spatial distribution of galaxies in these density bins is shown in Fig.~\ref{fig:cone_dens} for each of the GAMA regions (G09, G12 and G15). A random sample of galaxies is plotted such that there is an equal number of points in each density bin, and within a constant thickness of $18.1\Mpcns$, therefore giving a clearer view of how the galaxies are distributed according to overdensity.

%TABLE
\begin{table*}
\centering
\begin{tabular}{c c c c c r r}
\hline\hline
Label  &\multicolumn{2}{|c|}{\dddp}&$f_{\rm \delta}$&$f_{\rm \delta}$&\multicolumn{2}{|c|}{$N_{\rm \delta, \rm DDP1} /10^{3}$}\\
       & min   & max               & GAMA    & Mock                   & GAMA    & Mock                        \\
\hline
\bf{d1}     & $\mathbf{-1.00}$ & $\mathbf{-0.75}$             & $\mathbf{0.259}$ & $\mathbf{0.226 \pm0.011}$       & $\mathbf{2.18}$  & $\mathbf{1.88 \pm0.13}$             \\
d2     & $-0.75$ & $-0.55$             & $0.109$ & $0.149 \pm0.012$       & $2.31$  & $3.30 \pm0.32$             \\
d3     & $-0.55$ & $-0.40$             & $0.087$ & $0.101 \pm0.016$       & $2.72$  & $3.52 \pm0.55$             \\
\bf{d4}     & $\mathbf{-0.40}$ &  $\mathbf{0.00}$             & $\mathbf{0.189}$ & $\mathbf{0.175 \pm0.004}$       & $\mathbf{9.48}$  & $\mathbf{9.77 \pm0.29}$             \\
d5     &  $0.00$ &  $0.70$             & $0.168$ & $0.169 \pm0.008$       & $16.1$  & $16.7 \pm1.02$           \\
\bf{d6}     &  $\mathbf{0.70}$ &  $\mathbf{1.60}$             & $\mathbf{0.106}$ & $\mathbf{0.099 \pm0.003}$       & $\mathbf{17.3}$  & $\mathbf{16.9 \pm0.80}$           \\
d7     &  $1.60$ &  $2.90$             & $0.057$ & $0.053 \pm0.002$       & $16.2$  & $15.5 \pm1.05$           \\
d8     &  $2.90$ &  $4.00$             & $0.016$ & $0.016 \pm0.001$       & $7.21$  & $7.49 \pm0.55$             \\
\bf{d9}     &  $\mathbf{4.00}$ & $\mathbf{\infty}$            & $\mathbf{0.010}$ & $\mathbf{0.012 \pm0.001}$       & $\mathbf{7.57}$  & $\mathbf{9.34 \pm0.72}$             \\
\hline
\end{tabular}
\caption {\small{Table of DDP1 overdensity bins, listing overdensity limits, effective volume fraction ($f_{\rm \delta}$) of each bin (Eqn.~\ref{eqn:eff_vol}), and number of galaxies in DDP1 redshift range for GAMA and the mock catalogues, where the scatter is calculated as the variation between the individual mock catalogues. Overdensity bins used for comparison of LFs are d1, d4, d6 and d9 (in bold). A visual representation of these is shown in Fig.~\ref{fig:cone_dens}.}}
\label{tab:density_bins}
\end{table*}

\subsection{Colour}
\label{section:colour}
Observed galaxy colour is a strong indication of star formation history~\citep{Mahajan2009,Maller2009,Wetzel2012}, but also depends on properties such as metallicity and gas content. In agreement with Fig. 2 of~\citet{Mahajan2009}, we find there is a clear correlation between colour as defined here, and specific star formation rate (as measured by~\citet{Gunawardhana2013} using $\rm{H_{\alp}}$ flux). However, significant scatter in the correlation suggests our measure of colour cannot be used as a direct indication of star formation. The correlation and scatter are consistent over all overdensities, and we therefore do not expect a colour definition that is more indicative of star-formation to have any significant qualitative impact on our results.

The galaxy sample is split by colour to test for any further environmental dependence of the LF. Galaxies colours are defined by the $g-r$ rest frame colour, that depends only on the \emph{r}-band and \emph{g}-band apparent magnitudes, and the individual k-corrections in the \emph{r}- and \emph{g}-bands. 

Galaxies are assumed to have no difference in luminosity evolution between the \emph{r}- and \emph{g}-bands when rest frame colours are calculated. SDSS model magnitudes are used as apparent magnitudes when calculating colours, following the procedure of~\citet{Loveday2012}. The sample is split between blue and red at $\col = 0.63$, resulting in a mean colour of $\langle g-r\rangle=0.47 (0.74)$ for blue(red) galaxies. The left panel of Fig.~\ref{fig:g_r_Mg} shows this divide in colour (dashed vertical line) and how it splits up the sample of galaxies in \col for different ranges of \Mr. The chosen splits in colour are motivated by the clear bimodality seen in Fig.~\ref{fig:g_r_Mg}. Any luminosity dependent bimodality is small enough to be ignored for this analysis. The sample is also divided into 10 colour bins, defined by every $10^{th}$ percentile of the DDP1 galaxy sample, to determine how the LF changes with environment for narrow splits in colour.

The colour split in the mock catalogues is set by preserving the same fraction of red and blue galaxies as in GAMA. This cut is consistent with a cut based on the bimodality of the colour distribution in the mock catalogues, but is about $0.10$ mag bluer than the corresponding cut in GAMA. This is a known limitation of the colour distribution in the~\citeauthor{Bower2006} model, however it is encouraging that despite this colour offset, the colour distributions are similar, barring a much stronger bimodality in the mock catalogues.

\subsection{Luminosity Function}
\label{section:LF}
The galaxy LF is measured for the galaxies in each overdensity bin. Here we use the step-wise maximum likelihood (SWML) estimator (Efstathiou, Ellis, Peterson 1988), that does not require the assumption of a functional form for the LF. The LF, $\phi(M)\,\rm d\it M$, estimated using this method is normalised using the number of galaxies ($N$) within the volume defined by the redshift limits ($z_{\rm 1}$ and $z_{\rm 2}$) of the galaxy sample, and the solid angle of the survey ($\Omega$):

\begin{equation}
N = \it \Omega \int_{z_{\rm 1}}^{z_{\rm 2}}\, \rm d\it z \frac{\rm d\it V}{\rm d\it z \rm d\it \Omega} \int_{\it M_{\rm faint(\it z)}}^{\it M_{\rm bright(\it z)}} \it \phi(M)\,\rm d\it M\ .
\end{equation}

To take into account the effective volume populated by an overdensity bin, the overdensity is measured as in \S\ref{section:dens_meas} but at positions distributed uniformly within the volume. The corresponding effective volume fraction is estimated as the fraction of points within overdensity bin $\delta$:

\begin{equation}
f_{\rm \delta} = \frac{N_{\rm r, \rm \delta}}{N_{\rm r}},
\label{eqn:eff_vol}
\end{equation}
where $N_{\rm r, \rm \delta}$ is the number of randoms with a specific overdensity, including those with completeness greater than the threshold defined above, and $N_{\rm r}$ is the total number of randoms spanning the entire DDP volume. Galaxies are weighted by $1/f_{\rm \delta}$ when measuring the LF to estimate their abundance. As discussed in \S\ref{section:dens_meas}, due to the definition of overdensity, DDP galaxies from a given density bin will, in effect, cover a slightly smaller volume of the survey than non-DDP galaxies. DDP galaxies are weighted by $1/f_{\rm \delta, \rm DDP}$, with

\begin{equation}
f_{\rm \delta, \rm DDP} = \frac{N_{\rm r,\rm \delta,\rm DDP}}{N_{\rm r}},
\end{equation}
where $N_{\rm r, \rm \delta, \rm DDP}$ is the number of randoms, treated as DDP galaxies (and therefore having adding one to their DDP count), within a given overdensity bin $\delta$. This chosen normalisation of the LF in each environment is such that the total LF is obtained by a weighted sum over each environment, with the weight inversely proportional to the volume covered by that environment. 

We do not correct the GAMA data for any global imaging incompleteness. We assume that the main effect is to globally change the normalisation in all density bins. See ~\citet{Loveday2012} and Loveday et al. (in prep) for more information.

\subsubsection{Schechter function fits}
The LF is often well described by a~\citet{Schechter1976} function, that expressed in units of absolute magnitude is given by:

\begin{equation}
\phi(M) = \frac{\ln 10}{2.5} \phi^* 10^{0.4(M^*-M)(1+\alpha)} \exp(-10^{0.4(M^*-M)}),
\end{equation}
The Schechter function is specified by \alp, \Ms~and \phis~describing, respectively, the power law slope of the faint end, the magnitude at which there is a break from the power law (the `knee' of the LF), and the normalisation of the LF. The values of these parameters that best fit the LF are found by minimising $\chi^2$ over a grid of values of \alp, \Ms~and \phis, using the errors described in \S\ref{section:LFerrs}. Due to the shape of the Schechter function, there are known degeneracies between \Ms, \alp~and \phis. Appendix~\ref{appendix:Ms_alp_deg} presents degeneracies in \alp~and \Ms~in more detail.

\subsubsection{LF errors}
\label{section:LFerrs}

Errors for the GAMA LFs are estimated using jackknife errors from 9 samples, obtained by splitting each of the GAMA regions into a further 3 samples. Errors estimated from the scatter between the mock catalogues provide a reliable estimate accounting for sample variance. Despite the advantage of using the variation between mock catalogue as errors, we use jackknife errors for the data for the following reasons. When measuring the LF for samples split by a property for which the mock catalogues and GAMA do not agree (e.g. colour, see Fig.~\ref{fig:g_r_Mg}), the variation in the mock catalogues does not faithfully describe the constraints on the GAMA LF. The mock catalogues do not probe the full range of apparent magnitudes provided by GAMA (due to an imposed bright limit of $m_{\rm r} = 15.0$). Nevertheless, comparing jackknife errors within a mock catalogue with the variation between mock catalogues, we find they are compatible to the level required in this work. The errors used for the mock galaxy LFs are calculated as the standard deviation from the combined mock catalogue. If fewer than 5 galaxies contribute to a LF bin (shown by an open circle), errors on it cannot be estimated reliably and it is ignored when fitting a Schechter function.

Similarly, the variation of the best fitting Schechter function parameters between the mock catalogues or jackknife samples provides reliable errors with which we can constrain scaling relations for the parameters with overdensity, and subsequently assess the significance of these scaling relations.

\begin{figure}
\centering
\includegraphics[width=88mm]{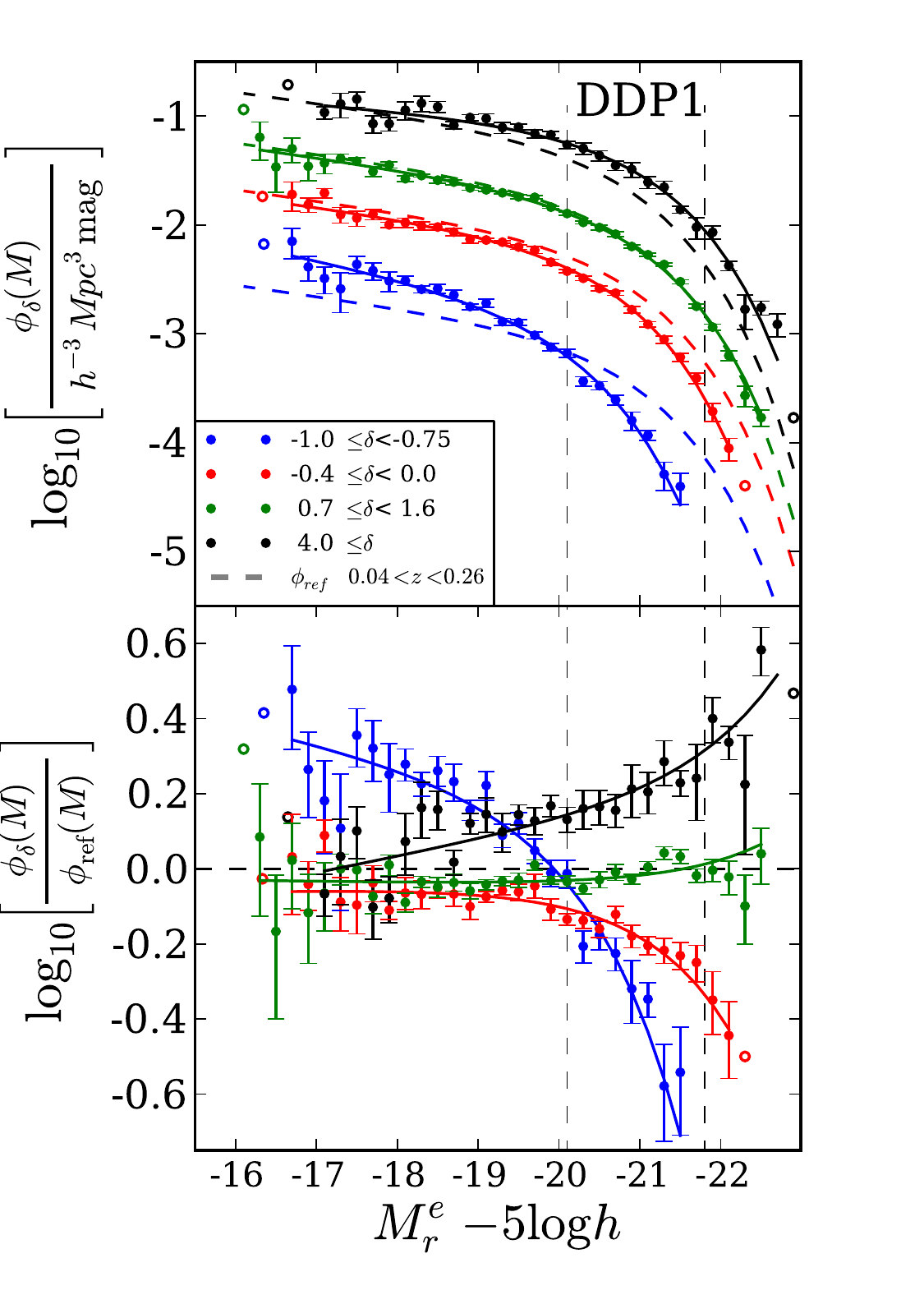}
\caption{\small{\emph{Top panel}: GAMA galaxy luminosity functions coloured according to environment (see key). The best fitting Schechter functions are shown by coloured solid lines, and the reference Schechter functions ($\phi_{\rm ref}$, see \S\ref{section:LF_d}) are given by dashed coloured lines (Eqn.~\ref{eqn:phi_ref}). \emph{Bottom panel}: ratio of the LF to the reference Schechter function, emphasizing the differences in shape between the LFs in different environments and the global LF. Errors in each panel are jackknife errors. Open circles are shown for LF bins where errors cannot be reliably estimated, these are not used when fitting a Schechter function. The dashed vertical lines show the absolute magnitude limits of the DDP sample.}}
\label{fig:LF_dens_DDPs_data}
\end{figure}

\begin{figure}
\centering
\includegraphics[width=88mm]{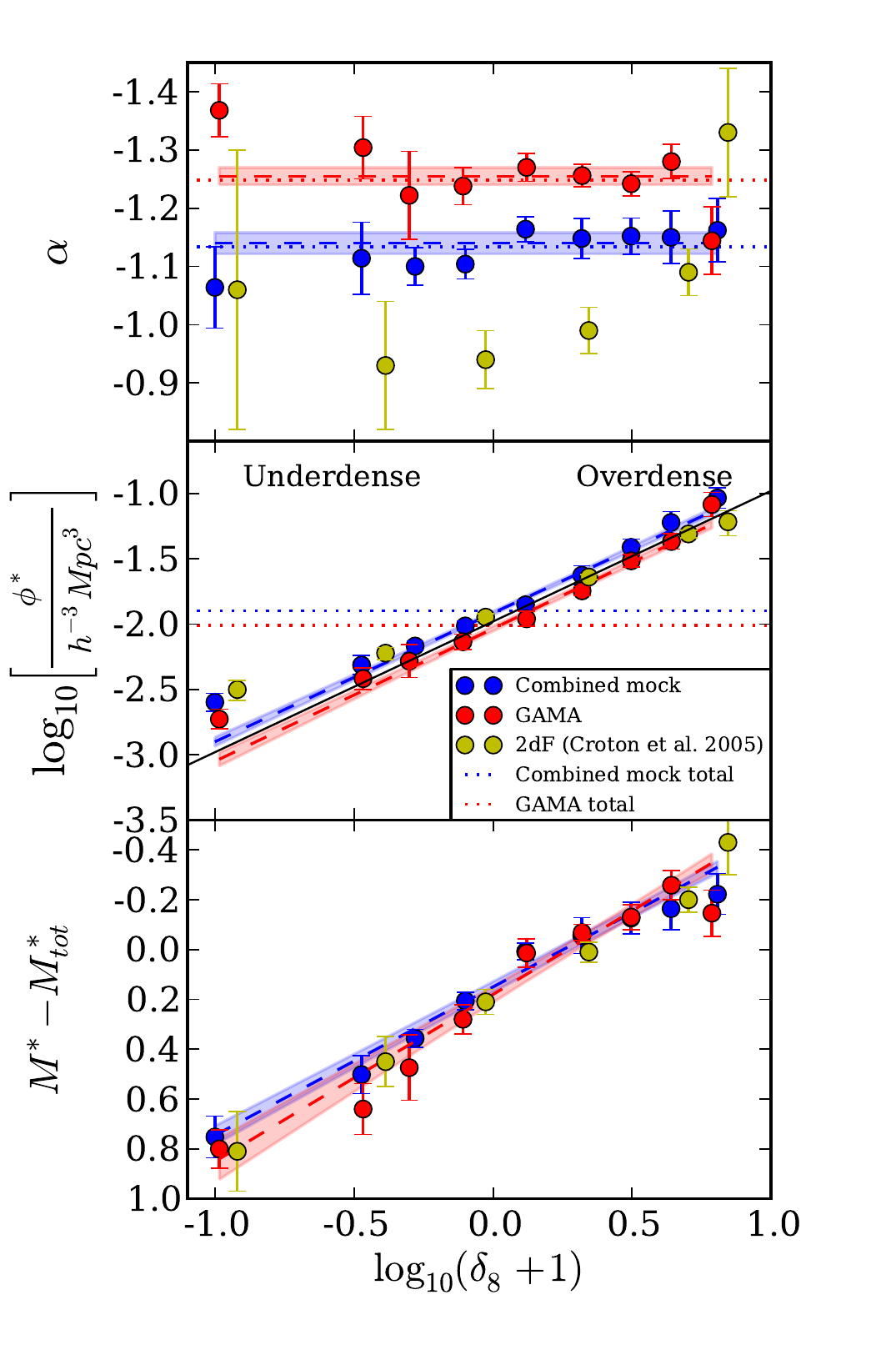}
\caption{\small{Schechter function parameters \alp~(\emph{top}), \phis~(\emph{middle}), and \Ms~(\emph{bottom}) as a function of environment for GAMA data (red) and simulated galaxy data (blue). \Ms~is plotted relative to $\Ms_{\rm tot}$, a reference value to compare different samples. $\alp_{\rm tot}$ and $\phis_{\rm tot}$, given by the reference Schechter function, are indicated by horizontal dotted lines for GAMA and the mock catalogues. Yellow points show the results of~\citet{Croton2005} from the 2dFGRS. Dashed lines show the best fitting relation as a function of overdensity, with the shaded regions indicating the uncertainty in the relations. \Ms~and $\log_{10}(\phis)$ vary linearly with \lonepd~ (the black solid line in the second panel indicates a gradient of unity), while \alp~seems to be broadly independent of overdensity.}}
\label{fig:schecht_dens_DDPs}
\end{figure}

\section{Results}
\label{section:Results}
We present LFs split by density in \S\ref{section:LF_d}, by redshift in \S\ref{section:LF_d_z} and by colour in \S\ref{section:LF_d_col}, to better understand any environmental, evolutionary and colour dependent trends.

\subsection{Environmental dependence of the LF}
\label{section:LF_d}
Overdensities are measured for all galaxies within the redshift limits of the DDP1 sample ($0.04<z<0.26$). Overdensity bins are listed in Table~\ref{tab:density_bins} for which galaxy LFs are measured. The top panel of Fig.~\ref{fig:LF_dens_DDPs_data} shows the LFs and best fitting Schechter function for 4 of these overdensity bins, from the most underdense (d1) to the most overdense bin (d9), with jackknife errors. As expected, these errors are smallest around the knee of the LF which is best constrained. 

Defining a \emph{reference} Schechter function allows us to compare how the shape of the LF varies with environment. Our reference Schechter function is based on the best fitting one to the LF of the full sample over all environments within the volume defined by the DDP1 sample ($\phi_{\rm tot}$), and is described by $\alp_{\rm tot}= -1.25$, $\Ms_{\rm tot}\Msu = -20.89$ and $\log_{10}{\phis_{\rm tot}/\phiu} = -2.01$ for GAMA.\footnote{The reference Schechter function for the mock galaxies is described by $\alp_{\rm tot}= -1.13$, $\Ms_{\rm tot}\Msu = -20.84$ and $\log_{10}{\phis_{\rm tot}/\phiu} = -1.90$.} These values are slightly different to those quoted in ~\citet{Loveday2012}. These differences are not of too much concern for this study, the reference function is derived using the same data and volume as that used here, thereby minimising any systematic effects introduced using slightly different data, volume or method of estimating the LF.

Assuming \phis~scales approximately with overdensity as $(1+\langle\dddp\rangle)$ (hereafter $1+\langle\dddp\rangle$ is noted as \onepdns), we scale our reference Schechter function for each density bin as

\begin{equation}
\phi_{\rm ref} = \frac{1+\dddp}{(1+\delta_{\rm tot})} \phi_{\rm tot}
\label{eqn:phi_ref}
\end{equation}
where $\phi_{\rm tot}$ is the Schechter function described above, and $\delta_{\rm tot}$ is the mean overdensity of the sample over the whole DDP volume, found to be $\delta_{\rm tot} = 0.007$.

 The dashed coloured lines in the top panel of Fig.~\ref{fig:LF_dens_DDPs_data} show the scaled reference Schechter function for each overdensity bin. We notice that our assumed scaling with $(\onepd)$ is a very good description of how \phis~scales with overdensity in all but the most extreme bins in overdensity. The deviation of the LFs in different environments from the scaled global LF is seen more distinctly in the lower panel of Fig.~\ref{fig:LF_dens_DDPs_data}. The variation seen at faint magnitudes indicates differences in the faint-end slope of the LF in different environments and those at bright magnitudes reflect a dependence of the characteristic luminosity on environment.

Fig.~\ref{fig:schecht_dens_DDPs} shows how the best fitting Schechter function parameters vary with \dddp for GAMA and the mock catalogues. \Ms~is shown as $\Ms - M^*_{\rm tot}$ with $M^*_{\rm tot}$ set by the reference Schechter function. Hence the variation of \Ms~with environment can be measured and compared to the $b_{\rm J}$-band results of~\citet{Croton2005} from 2dFGRS.
 We note that the best fitting Schechter function for the total GAMA sample within the DDP redshift limits (defined above) is in very good agreement with that found in the mock catalogues.

 The uncertainty on the Schechter parameters correlates strongly with sample size, indicated in Table~\ref{tab:density_bins}. This mostly explains the observed bin to bin variations of the errors. The strong correlations between \alp, \Ms~and \phis~also have an effect on the inferred errors. A full covariance matrix analysis would be required in order to statistically constrain these correlations, but degeneracies between \Ms~and \alp~are clearly shown in Appendix~\ref{appendix:Ms_alp_deg}, and can be ruled out with high confidence to be the cause of the trends with overdensity.

The coloured dashed lines in Fig.~\ref{fig:schecht_dens_DDPs} show how the Schechter function parameters scale with overdensity. The variation in the scaling relations due to sample variance (as indicated by the shaded regions) is found by calculating the scatter between the best fitting lines for each jackknife sample or mock catalogue. Table~\ref{tab:best_fit_d_c} gives parameters for the linear fits, shown by the dashed lines. \alp does not show any specific trend with environment and we therefore fit it as a constant. \Ms~and \phis~vary significantly with environment. This is expected for \phis, since the most overdense regions have the highest number density of galaxies. 

\Ms~brightens linearly with \lonepd, at very similar rates for GAMA and the mock catalogues. This is characterised by a negative slope, given in Table~\ref{tab:best_fit_d_c}. 

\begin{figure*}
\centering
\includegraphics[width=176mm]{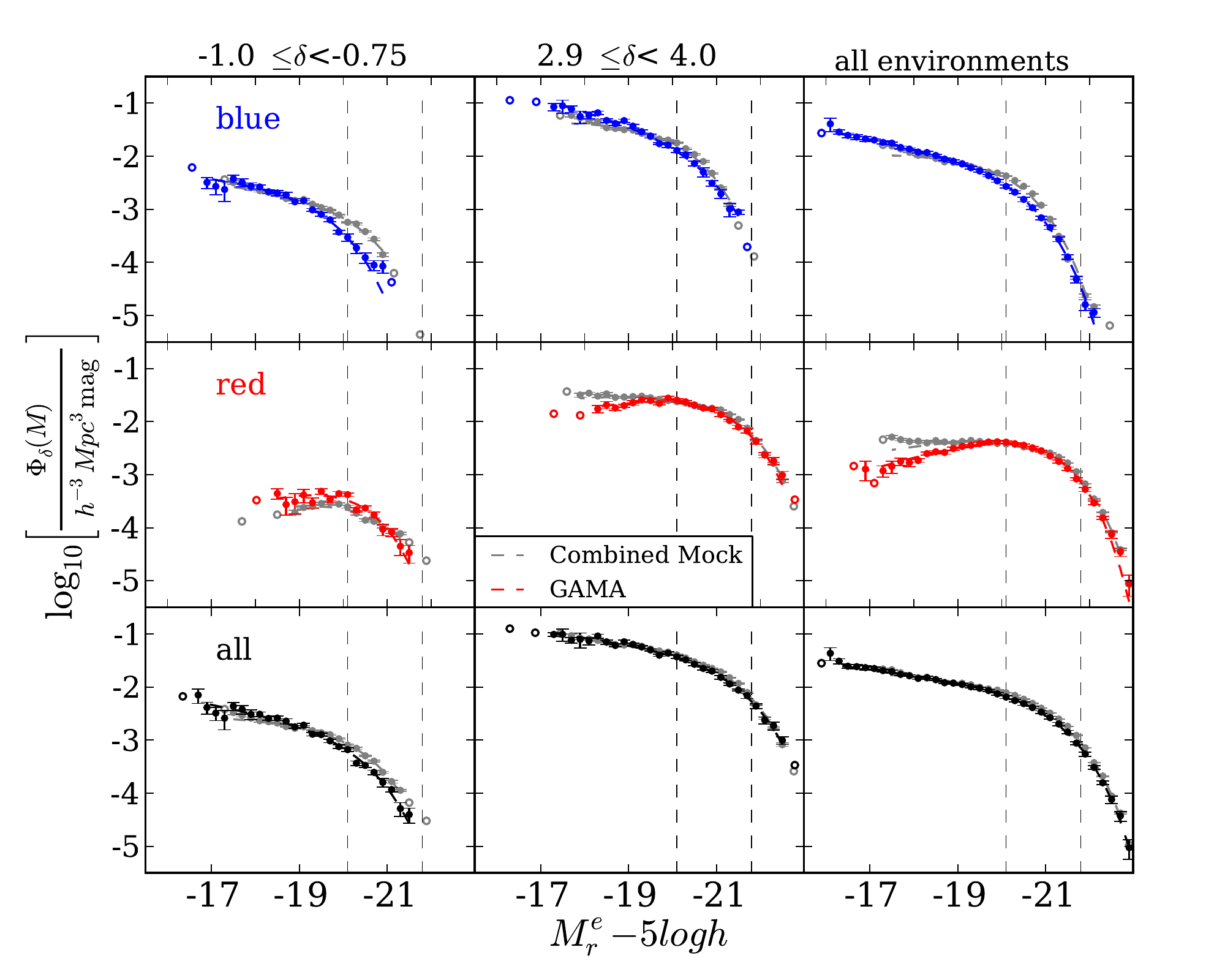}
\caption{\small{Luminosity functions for mock galaxies (grey) compared to GAMA galaxies, for different splits in colour (top to bottom) and overdensity (left to right). From left to right: LFs in the most underdense environment, an overdense environment and the global LFs (i.e. not split by density). Top to bottom: LFs for blue, red and all galaxies. Open circles are shown for LF bins where errors cannot be reliably estimated, these are not used when fitting a Schechter function. The LFs are remarkably similar between the mock catalogues and GAMA, given that only the total LF (bottom right) has been constrained in the mock catalogues. The more significant discrepancies between GAMA and the mock catalogues are at the bright end of the blue LFs, and the faint end of the red LFs (see \S\ref{section:phys_interp} for further discussion).}}
\label{fig:LFs_gam_mock}
\end{figure*}

The bottom panels of Fig.~\ref{fig:LFs_gam_mock} show how the LFs for the GAMA and the combined mock catalogue compare in the most underdense bin (d1), an overdense bin (d8) and for the total sample. The GAMA and mock galaxy LFs are very similar in the two extreme environments. 

The results found from GAMA are mostly in good agreement with those from~\citet{Croton2005}, although the values of \alp~in different environments seem somewhat inconsistent, as discussed further in \S\ref{section:quant_disc}.

\begin{figure*}
\centering
\includegraphics[width=170mm]{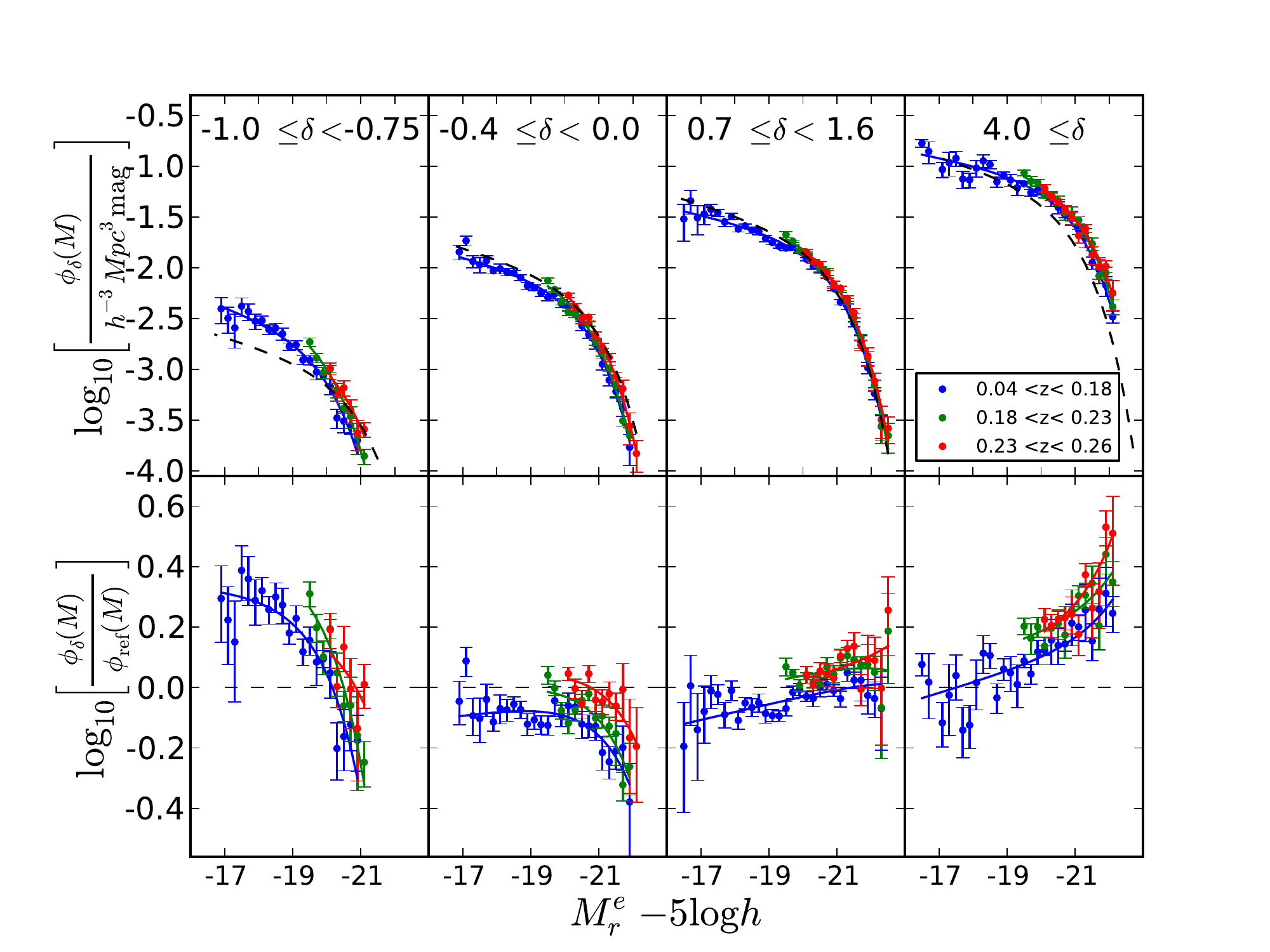}
\caption{\small{\emph{Top panel}: GAMA LFs for 4 different overdensity bins (same as in Fig.~\ref{fig:LF_dens_DDPs_data}), from most underdense (left) to most overdense (right), split by redshift (see key). The solid coloured curves show the best fitting Schechter functions, and the black dashed curves show the reference Schechter function ($\phi_{\rm ref}$, see \S\ref{section:LF_d}) for the whole redshift range (as in Fig.~\ref{fig:LF_dens_DDPs_data}. \emph{Bottom panel}: ratio of the LFs to the reference Schechter function. Errors in each panel are jackknife errors.}}
\label{fig:LF_dens_z_data}
\end{figure*}

\begin{figure*}
\centering
\includegraphics[width=176mm]{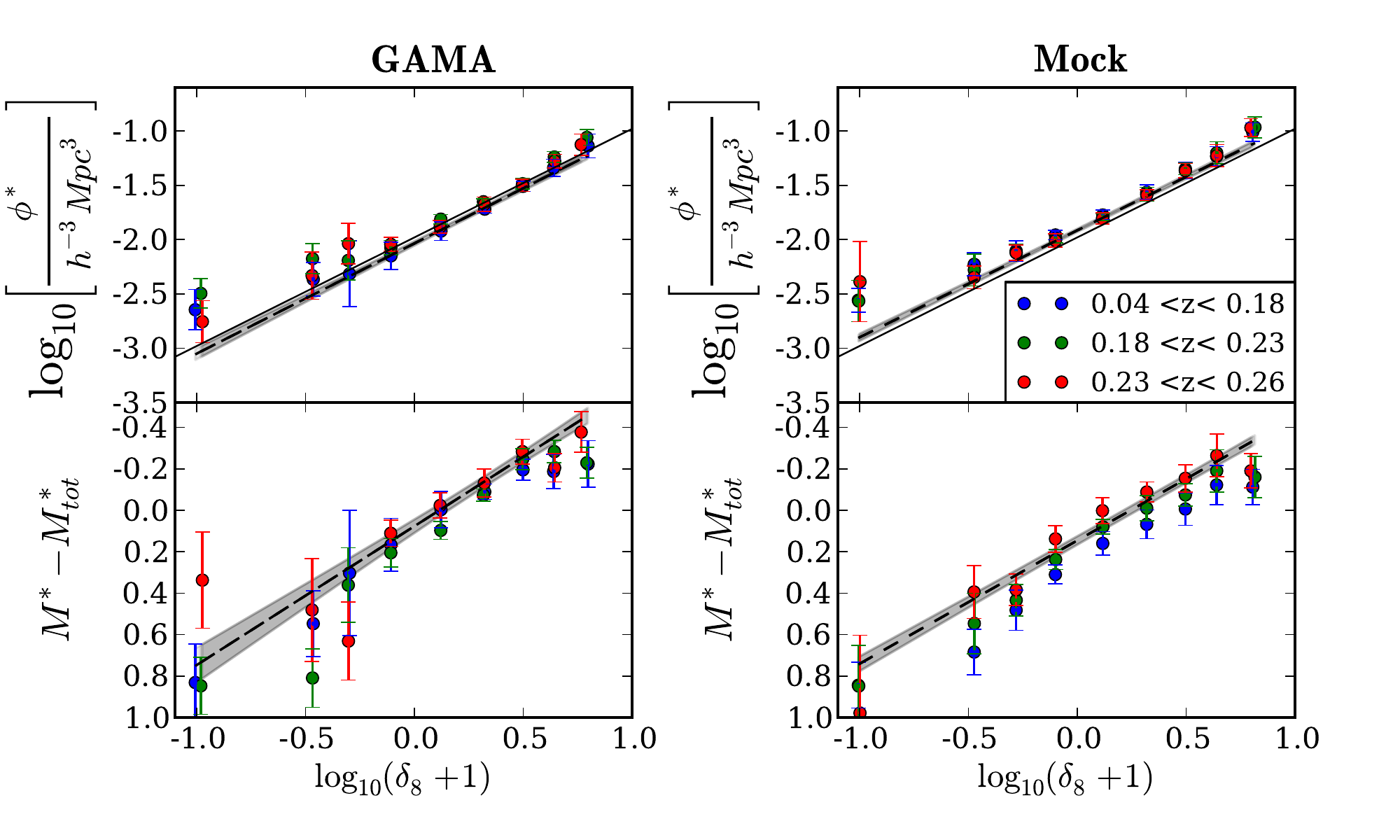}
\caption{\small{Best fitting Schechter function parameters \phis~and \Ms~as a function of overdensity for GAMA (\emph{left}) and the mock catalogues (\emph{right}) coloured according to redshift (see key). Uncertainties are jackknife errors (for GAMA) or scatter in the mock catalogues (for combined mock catalogue). The scalings of \phis~and \Ms~with overdensity for the total sample not split by redshift are shown by dashed lines and shaded regions. The black solid lines in the upper panels indicate a gradient of unity.}}
\label{fig:schecht_dens_z_data}
\end{figure*}

\subsection{Evolution of the LF dependence on environment}
\label{section:LF_d_z}

To determine whether or not the dependence of the LF on environment evolves with redshift, we measure the LF for the same environments given above, but for 3 separate redshift slices of equal volume: $0.04<z<0.18$, $0.18<z<0.23$ and $0.23<z<0.26$. The highest redshift sample only probes galaxies brighter than $\Mr=-19.8$, resulting in the faint end of the LF being poorly constrained. Therefore, when fitting Schechter functions in the two higher redshift slices, \alp is fixed to the best fitting value of the lowest redshift slice in each environment, and only \Ms~and \phis~are treated as free parameters. This value of \alp~is highly consistent with that measured over the whole redshift range, only deviating by at most $\pm0.02$. To constrain any evolution in \alp, a deeper survey is necessary, allowing the LF to be constrained down to lower luminosities at higher redshifts. The resulting LFs are shown in Fig.~\ref{fig:LF_dens_z_data}.

Fig.~\ref{fig:LF_dens_z_data} shows a small offset in the LFs between different redshifts for underdense environments. These offsets can be accounted for by a small density evolution, that has not been taken into account in this analysis, and/or an additional luminosity evolution (see \S\ref{subsection:GAMA}). These are very degenerate and cannot be constrained well enough through this analysis due to the sample size considered, but since this trend is visible in all 3 GAMA regions, it is evident that there is some small density and/or additional luminosity evolution in the LF, especially in underdense environments. Fig.~\ref{fig:schecht_dens_z_data} however shows that sample variance within GAMA is larger than this offset.

The best fitting values for \Ms~and \phis~as a function of overdensity are shown in Fig.~\ref{fig:schecht_dens_z_data} for GAMA and the mock catalogues (left and right panels respectively). The dashed coloured lines show the linear fits to the total samples split by overdensity, as shown in Fig.~\ref{fig:schecht_dens_DDPs}. Although the best fitting values for \phis~and \Ms~for different redshifts do not closely follow the scaling relation with overdensity of the total sample, the degeneracies in \phis~and \Ms~are likely to affect these results such that a value for \Ms~that is measured to be ``too faint'' according to the scaling, can have a good fit in conjunction with ``too high'' a value for \phis. The evolution of the two parameters is not apparent in Fig.~\ref{fig:schecht_dens_z_data} over the luminosity evolution already accounted for.

\begin{figure*}
\centering
\includegraphics[width=176mm]{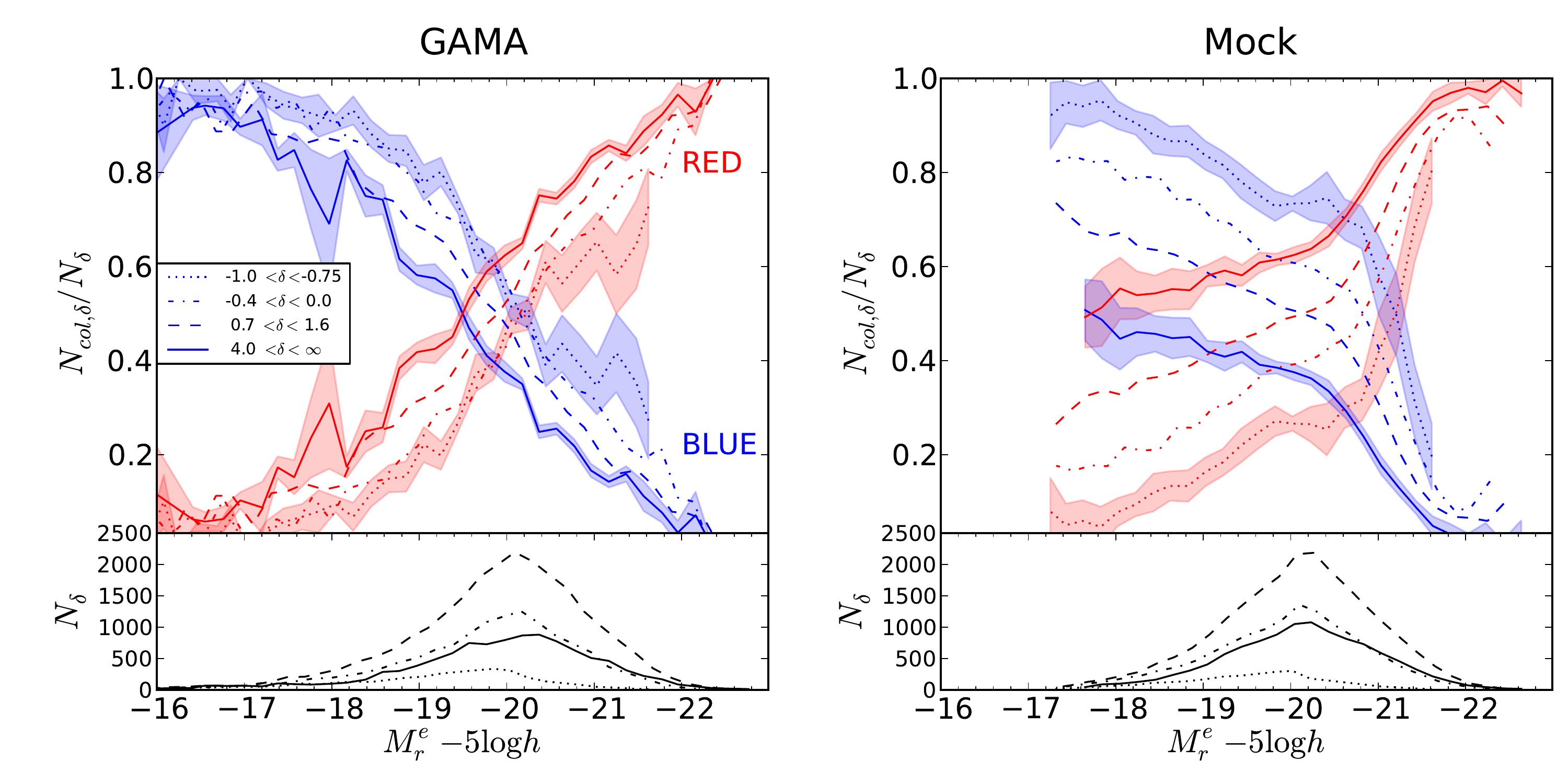}
\caption{\small{\emph{Top panels}: Red and blue galaxy fractions for 4 environments (see key) as a function of absolute magnitude, for GAMA (\emph{left}) and the mock catalogues (\emph{right}). The shaded regions in the right panel show the scatter from individual mock catalogues and in the left panels show jackknife errors in GAMA for the most overdense and most underdense bins. Lines are coloured according to galaxy colour. The fraction of red galaxies increases with overdensity and brightness, whereas the fraction of blue galaxies decreases with increasing overdensity and brightness. \emph{Bottom panels}: Distribution of absolute magnitudes for the overdensity bins shown in the top panel. While presenting similar overall trends, the mock catalogues have a significantly different distribution of colour fractions to GAMA. This is discussed in \S\ref{section:phys_interp}.}}
\label{fig:col_fract}
\end{figure*}

\begin{figure*}
\centering
\includegraphics[width=176mm]{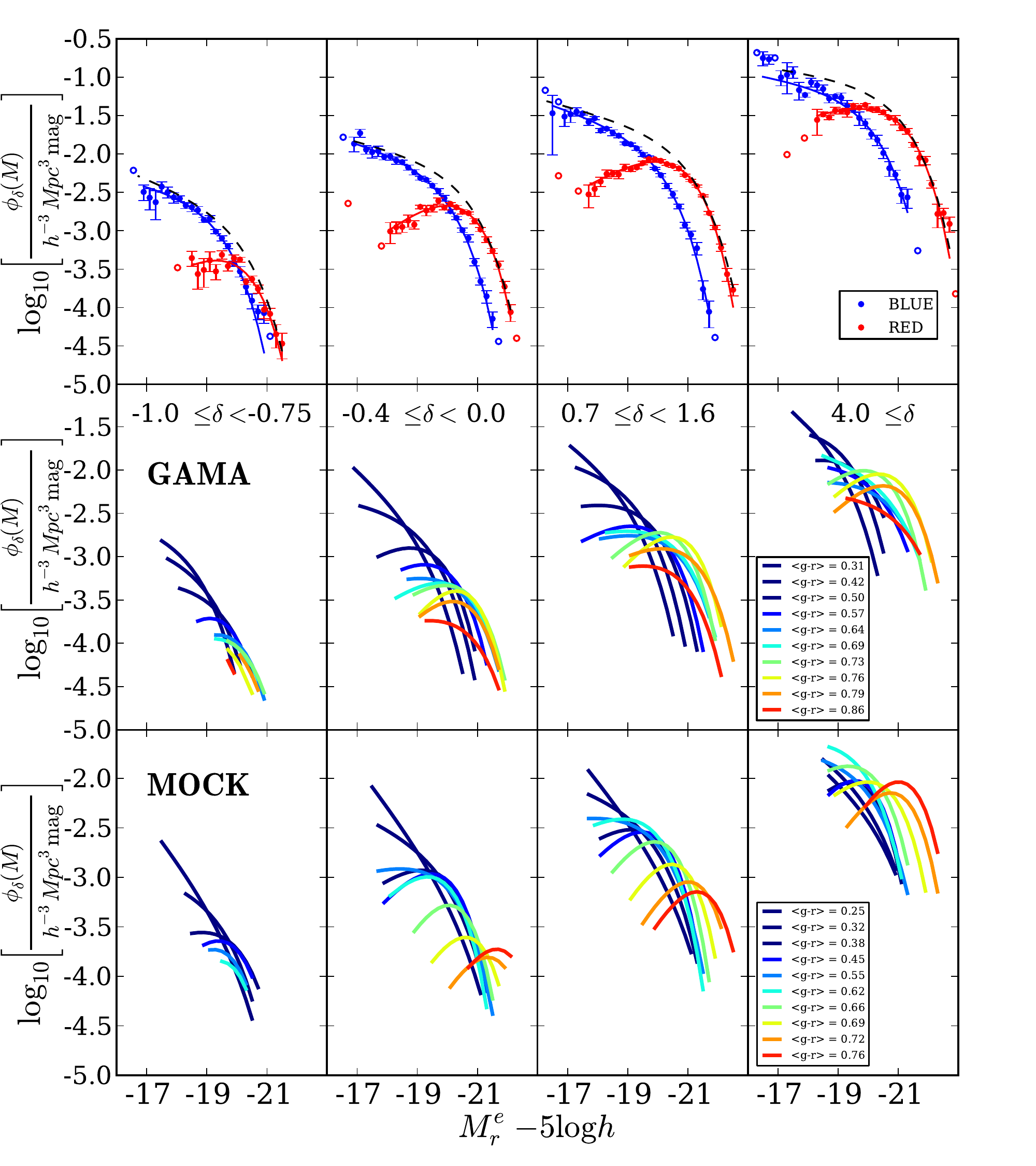}
\caption{\small{\emph{Top}: GAMA LFs and best fitting Schechter functions for red and blue galaxies, split by environment as indicated in the central panels. The dashed lines show the total Schechter function in each overdensity bin (as in Fig.~\ref{fig:LF_dens_DDPs_data}). Open circles are shown where LF errors cannot be reliably estimated. \emph{Middle} : Schechter function fits as a function of colour, from the bluest to the reddest galaxies in 10 narrow colour bins (see Fig.~\ref{fig:g_r_Mg}). The shape of the LF depends strongly on colour, and the transition between the shapes of the blue and red LFs is clear. Schechter functions are not extrapolated beyond the range of the measured LF in each colour bin. \emph{Bottom} : The same as the middle panels but for the mock catalogues. The mock catalogues show the same general trend from red to blue as GAMA, but in detail show some clear differences for the LFs measured for samples defined by narrow bins in colour.}}
\label{fig:LF_dens_c_data}
\end{figure*}

\begin{figure*}
\centering
\includegraphics[width=176mm]{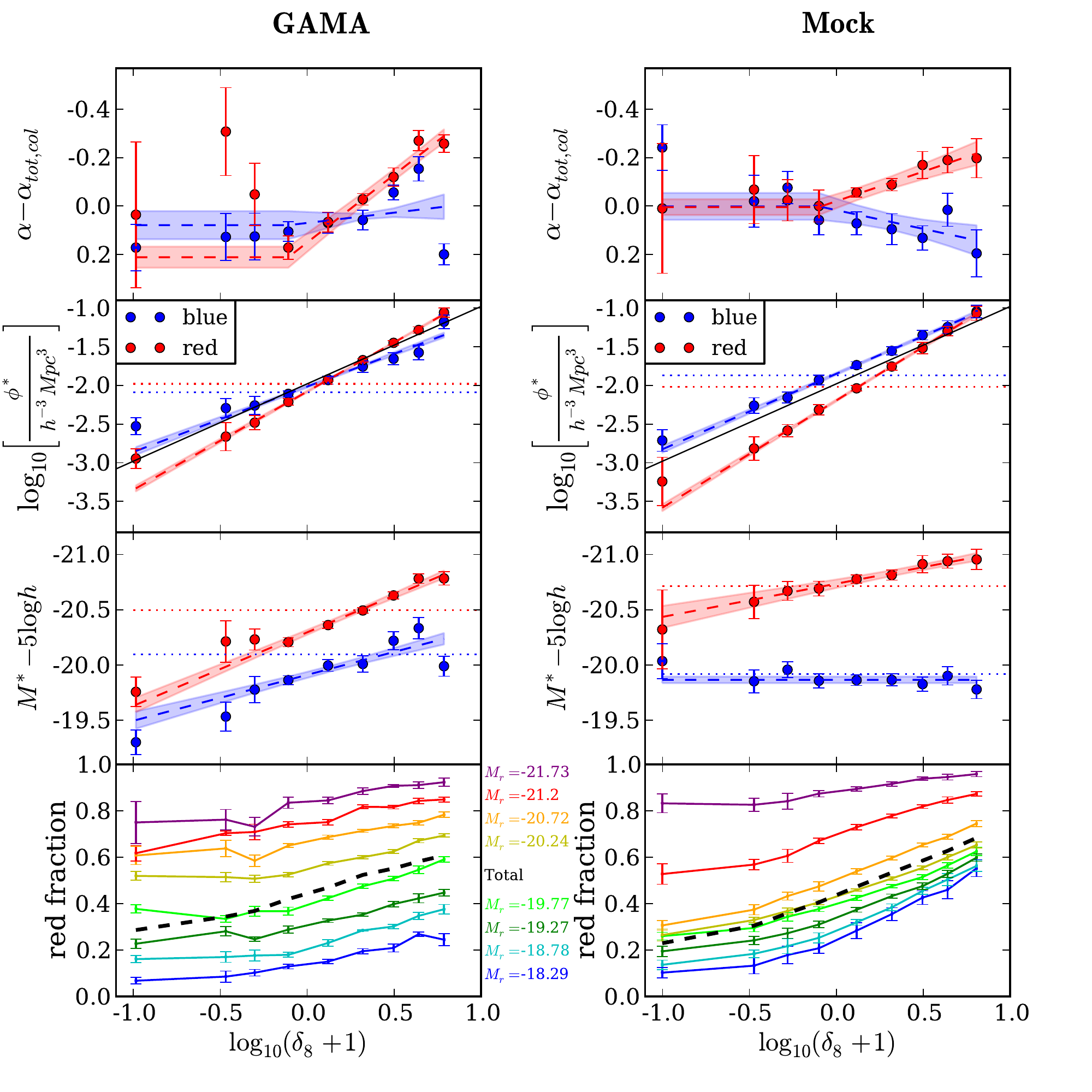}
\caption{\small{\emph{Top 3 panels}: Schechter function parameters as a function of overdensity for blue and red galaxies in GAMA (\emph{left}) and mock catalogues (\emph{right}). \alp~is plotted with respect to the reference Schechter function for each colour ($\alp_{\rm tot, \rm col}$, see \S\ref{section:LF_d_col} for values). The dotted lines show the Schechter function parameters for the samples not split by environment. As in Fig.~\ref{fig:schecht_dens_DDPs}, shaded regions show the uncertainty in the line fits, and the black solid lines shows a gradient of unity. \emph{Bottom}: Fraction of galaxies classified as red as function of overdensity for 8 bins in absolute magnitude. Labels shown are the median absolute magnitudes in each bin. Uncertainties shown are jackknife errors (left) or scatter in the mock catalogues (right). The red fraction for the total sample is given by the black dashed line.}}
\label{fig:schecht_dens_c_data}
\end{figure*}

\begin{table*}
\centering
\begin{tabular}{l l r r r r}
\hline\hline
 colour& Schechter Parameter   &     \multicolumn{2}{|c|}{GAMA}                &         \multicolumn{2}{|c|}{Mocks}           \\
       &                       & $a_0$\quad\quad\quad  & $a_1$\quad\quad\quad  &  $a_0$\quad\quad\quad  & $a_1$\quad\quad\quad  \\
\hline
 all   & \alp              & $-1.25\pm0.01$        & - \quad\quad\quad     & $-1.14\pm0.01$         & - \quad\quad\quad     \\
       & $\log_{10}\phis$     & $-2.03\pm0.03$        & $1.01\pm0.06$         & $-1.92\pm0.02$         & $0.98\pm0.05$         \\
       & $\Ms\Msuns$           & $-20.70\pm0.03$       & $-0.67\pm0.07$        & $-20.69\pm0.02$        & $-0.60\pm0.06$        \\
 blue  & \alp              & $-1.30\pm0.01$        & $-0.08\pm0.01$        & $-0.95\pm0.01$         & $0.15\pm0.01$         \\
       & $\log_{10}\phis$     & $-2.01\pm0.02$        & $0.85\pm0.07$         & $-1.85\pm0.03$         & $0.97\pm0.07$         \\
       & $\Ms\Msuns$           & $ -19.91\pm0.03$      & $-0.42\pm0.08$        & $-19.87\pm0.02$        & $-0.00\pm0.03$        \\
 red   & \alp              & $-0.23\pm0.12$        & $-0.56\pm0.25$        & $-0.67\pm0.04$         & $-0.25\pm0.12$        \\ 
       & $\log_{10}\phis$     & $-2.08\pm0.02$        & $1.27\pm0.05$         & $-2.19\pm0.03$         & $1.38\pm0.07$         \\  
       & $\Ms\Msuns$           & $-20.30\pm0.02$       & $-0.67\pm0.06$        & $-20.74\pm0.03$        & $-0.30\pm0.07$        \\
\hline
\end{tabular}
\caption {\small{Table of coefficients for best fitting relations describing how the Schechter function parameters vary with overdensity for all, red and blue galaxies, as shown in Fig.~\ref{fig:schecht_dens_DDPs} and Fig.~\ref{fig:schecht_dens_c_data} for GAMA and the mock catalogues. Scaling coefficients are given for $Y = a_0 + a_1\log_{10}(\onepdns)$ where $Y=\log_{10}{\phis/\phiuns}$ or $Y=\Ms\Msuns$. \alp~(all) is fit by $a_0$, while \alp~(colours) is fit by the relation given in Eqn.~\ref{eqn:alp_dens}. Statistical errors from the jackknife resamplings (data) or variations in the mock catalogues (mocks) are given.}}
\label{tab:best_fit_d_c}
\end{table*}

\subsection{Dependence of the Luminosity Function on Environment and Colour}
\label{section:LF_d_col}
To determine whether or not there is any environmental dependence of the LF over any colour-density relation, we look at how the LF varies for blue and red galaxies as a function of overdensity. The mock galaxy LFs can then be compared to the GAMA LFs to determine where the galaxy formation models do not agree with GAMA. 

It can clearly be seen from Fig.~\ref{fig:LFs_gam_mock} that although remarkably similar, the shapes of the LFs for the mock galaxies do not entirely agree with the shapes of the GAMA LFs when split by colour. The total \emph{r-}band LF for the mock galaxies matches the GAMA \emph{r-}band LF by construction, thus the bottom right panel shows very good agreement between GAMA and the mock galaxies. However, when splitting the LFs by density and colour, it is clear that the mock catalogues predict too many bright blue galaxies in underdense environments. Similarly too few faint red galaxies are predicted by the mock catalogues in underdense regions, but too many faint red galaxies are predicted in overdense regions. The faint end of the blue LF in underdense environments and the bright end of the red LF in overdense environments agree very well with the GAMA LF. Fig.~\ref{fig:col_fract} shows that blue galaxies tend to dominate underdense and red dominate overdense environments, these are therefore most influential in determining the LF over all environments, as seen in the right hand panels of Fig.~\ref{fig:LFs_gam_mock}.

The LFs split by red and blue galaxies for 4 different environments in GAMA are shown in the top panels of Fig.~\ref{fig:LF_dens_c_data}. The shape of the LF clearly differs between red and blue galaxies~\citep{Loveday2012,DePropris2013}, but it is not obvious that the shape of LFs for blue and red populations vary with environment. This can be investigated further by looking at the shape of the LF for 10 narrow splits in colour, representing 10 percentile intervals in the colour distribution (see Fig.~\ref{fig:g_r_Mg}). The LFs for these splits are shown in the middle (bottom) panels of Fig.~\ref{fig:LF_dens_c_data} for GAMA (mock catalogues).

The shape of the LF for any given narrow range of colour can be seen to vary with increasing density. In particular, the LF of the extreme blue sample does not seem to vary significantly with density, while the faint-end slope of the LF for redder samples tends to become steeper with overdensity.

In Fig.~\ref{fig:LF_dens_c_data}, the mock galaxy LFs brighten as the sample gets redder, and the number of faint galaxies at a fixed luminosity decreases. Similar trends are seen in GAMA, where generally redder samples tend to contain brighter galaxies, but the variation between the LFs of the reddest samples is much smaller than is predicted by the mock catalogues. Although red galaxies clearly dominate the most overdense regions at bright luminosities, Fig.~\ref{fig:LF_dens_c_data} suggests that this increase in the number of red galaxies with overdensity is mainly caused by the intermediate red population rather than the very reddest. 

The Schechter function parameters \alp, \Ms~and \phis~for the GAMA LFs are shown in the left panel of Fig.~\ref{fig:schecht_dens_c_data}: $\alp\:$ is shown with respect to $\alp_{\rm tot,\rm col}$, the faint-end slope of the total LF for each colour sample. This allows the variation of \alp~with overdensity to be compared between different colour samples, especially as the values of $\alp_{\rm tot, \rm col}$ for GAMA and the mock catalogues are different between red samples (GAMA: $\alp_{\rm tot, \rm red} = -0.38$, mock catalogues: $\alp_{\rm tot, \rm red} = -0.65$) and blue samples (GAMA: $\alp_{\rm tot,\rm blue} =-1.37$, mock catalogues: $\alp_{\rm tot, \rm blue} = -0.96$).

Both red and blue galaxy samples display linear dependencies of \phis~and \Ms~with \lonepd. The best fitting parameters describing these dependencies are given in Table~\ref{tab:best_fit_d_c}. \alp~appears to follow a relation of the form:

\begin{equation}
\label{eqn:alp_dens}
\alpha = \left\{ \begin{array}{ll}
  a_0 &\mbox{ $\delta_{\rm 8} \le -0.2$}\\
  a_0 + a_1\log_{10}{(1+\delta_8)} &\mbox{ $\delta_{\rm 8} > -0.2$},
\end{array} \right.
\end{equation}

This implies that the faint end of the LF steepens with overdensity only in overdense regions for a given galaxy population. 
\phis~increases at a significantly faster rate with overdensity for red galaxies than for blue galaxies, which is consistent with blue galaxies dominating underdense regions and red galaxies dominating overdense regions. The value of \phis~for red and blue samples with overdensities around $\dddp=0$ is similar, suggesting a similar fraction of red and blue galaxies populate average density environments.

The $3^{\rm rd}$ panel down on the left in Fig.~\ref{fig:schecht_dens_c_data} shows that \Ms~brightens at a faster rate with overdensity for blue galaxies than for red galaxies in GAMA. In underdense regions, the offset between \Ms~for the two colour sub-samples is as small as $\sim0.1$ mag, whereas in the most overdense regions there difference becomes as large as $\sim0.5$ mag.  The significant offset ($\sim0.45$ mag) between $\Ms_{\rm tot}$ for blue and red galaxies (shown by the dotted horizontal lines), can be understood from the change in \phis~with environment: \Ms~in overdense regions is determined by red galaxies, whereas in underdense regions it is determined by blue galaxies.

The changes in best fitting Schechter function parameters with environment for the mock catalogues are qualitatively similar to the observational data (see right panels of Fig.~\ref{fig:schecht_dens_c_data}). $\alp\:$ shows a slightly different trend to that observed in GAMA. While the faint-end slope appears to steepen with environment in GAMA (more so for red galaxies than for blue), the faint-end slope for blue galaxies in the mock catalogues tends to become shallower for more overdense environments.

The variation in the amount of blue and red galaxies with overdensity predicted by the mock catalogues is as significant as that observed in GAMA ($2^{\rm nd}$ panels down in Fig.~\ref{fig:schecht_dens_c_data}), although the predicted number of blue galaxies at higher overdensities is slightly higher than is observed. The variation in \Ms~with environment for colour sub-samples predicted by the mock catalogues is inconsistent with GAMA. Although the mock catalogues correctly predict red galaxies brightening with overdensity, there is no dependence of \Ms~on environment predicted for blue galaxies, while \Ms~for red galaxies shows a weaker brightening with overdensity than is observed, causing \Ms~to be predicted too bright in the most underdense environments.

The fraction of red galaxies as a function of overdensity for bins in absolute magnitude is shown in the lower panel of Fig.~\ref{fig:schecht_dens_c_data}, where as expected we find that brighter samples have a consistently higher red fraction than fainter samples, and that the fraction of red galaxies increases with overdensity for all luminosities. The mocks (right panel) show that although qualitatively similar, there are some differences in the red fraction of the bright magnitude bins (except the very brightest bins) for the most underdense environments, and in the faintest magnitude bins for the most overdense environments.

\section{Discussion}
\label{section:Disc}

We have used GAMA to measure luminosity functions for different environments, redshifts and galaxy colours. Here we summarise our findings and discuss the implications for galaxy formation.

\subsection{Quantitative Description}
\label{section:quant_disc}
A density defining population (DDP) of galaxies is used as a tracer of the underlying matter distribution. It provides a means by which to measure how the properties of the galaxy population, such as luminosity and colour, vary with environment. There is generally a good agreement between different DDP tracers used to measure overdensity, as discussed in Appendix~\ref{appendix:DDP}. Mapping the most extreme environments is sensitive to the choice of DDP tracer, and so mock galaxy catalogues constructed from simulated galaxy data are required for quantitative comparisons to models of galaxy formation.

GAMA is a deeper (up to 2 mags) and more spectroscopically complete survey than those that have previously been used to investigate the variation in the galaxy LF with environment (2dFGRS, SDSS). Hence it provides more reliable environment measures over a large range of environments. 

The galaxy LF is measured in 9 overdensity bins from GAMA, over the redshift range of $0.04<z<0.26$. The LFs for 4 of these density bins are shown in Fig.~\ref{fig:LF_dens_DDPs_data}. The shape of the LF is found to vary smoothly with overdensity, with little change in the faint-end slope \alp, but where the characteristic magnitude \Ms~and characteristic number density $\log_{10}{\phis}$ vary linearly with \lonepd, as can be seen in Fig.~\ref{fig:schecht_dens_DDPs}. Although a Schechter function is a poor fit to the total galaxy sample, it is a reasonable description in underdense regions. 

Assuming galaxy overdensity relates to mass overdensity as $\delta_{\rm g} = b_{\rm g}\delta_{\rm mass}$, like in a linear bias model, and that \phis~varies with mass overdensity as $\phis = (1+\delta_{\rm mass})$, we expect a linear relation between $\log_{10}{\phis}$ and \lonepd~through our chosen method of measuring overdensity, the slope of which is $1/b_{\rm g}$. We find a slope of $\log_{10}{\phis}$ with \lonepd~of $1.01\pm0.06$, consistent with a galaxy bias of $b_{\rm g} = 0.99$. This is slightly higher than $b_{\rm g}=1.20$ measured by~\citet{Zehavi2011} for the absolute magnitude range of our DDP sample. This approximation for the scaling of \phis~is only valid for $\dddp\ll1$, and so we do not expect this scaling to work for our most overdense bins. If only considering the 5 lowest density bins (lower than e.g. $\lonepd = 0.3$, corresponding to the density beyond which our approximation is invalid), we find a slope of $0.87\pm0.09$, consistent with the bias measured by \citet{Zehavi2011}. Measuring the variation of the normalisation of the luminosity function in underdense regions with different DDP galaxies could be a way to measure the bias of galaxies. However due to the small range of overdensities for which the approximation works, a much larger galaxy sample is needed to actually measure the linear galaxy bias.

The degeneracies between \alp, \Ms~and \phis~affect our ability to constrain the shape of the LF. These degeneracies have an impact on the best fitting Schechter functions for each jackknife sample or for individual mock catalogues (see Appendix ~\ref{appendix:Ms_alp_deg}), resulting in large uncertainties on these parameters. When using a larger sample over a large volume in the survey (e.g. the $5^{\rm th}$ density bin), degeneracies are more easily overcome by the ability to better constrain one parameter (\phis). Appendix~\ref{appendix:Ms_alp_deg} shows that the variation of each parameter with overdensity is more significant than these degeneracies.

Comparing our results for the galaxy population as a whole to those of~\citet{Croton2005}, we find agreement that the galaxy LF varies smoothly with environment. The faint-end slope \alp~does not show any significant variation with environment, suggesting the abundance of faint galaxies varies linearly with overdensity as \phis~only. This suggests that the physical process involved in suppressing the formation of faint galaxies is likely to be an internal process, such as supernovae or photo-ionisation, rather than an environmental one. From Fig.~\ref{fig:schecht_dens_DDPs} it is clear that the values of \alp~presented by~\citet{Croton2005} are much shallower (by up to $\Delta\alp\sim0.3$) than those found for GAMA. The extra depth gained when using GAMA data allows the LF to be measured over a larger magnitude range $4.65>M_{\rm r}-\Ms>-2.35$, which is 2 mags fainter than~\citet{Croton2005} ($2.65>M_{\rm b_{\rm J}}-Ms>-2.35$), providing the ability to better constrain the faint end of the LF using GAMA.

 Our conclusion that \Ms~varies linearly with \lonepd~is similar to the 2dFGRS results of~\citet{Croton2005}. However, we find a slightly stronger dependence of \Ms~on overdensity. The 2dFGRS is selected in the $\emph{b}_{\rm J}$-band, and the sample contains a predominantly blue population of galaxies compared with our r-band selected analysis. Fig.~\ref{fig:schecht_dens_c_data} shows clearly that blue galaxies have a much slower increase in \phis~with overdensity than red galaxies, and a fainter \Ms~in all environments. Thus when considering the whole sample, a smaller fraction of red galaxies in overdense environments will cause less brightening of \Ms~with overdensity. This highlights the importance of considering the galaxy population used when analysing the shape of the LF.

These results are also consistent with those presented in Figs. 11 and 12 of~\citet{Verdes-Montenegro2005}, who collate previous estimates of the LF for different environments and surveys and compare how \Ms~and \alp~vary as a function of density, finding a brightening of \Ms~with environment density, and only a weak steepening of the faint end.

The brightening of \Ms~in denser environments suggests physical processes which either suppress the bright end of the LF in more underdense environments or induce a brightening of galaxies in overdense environments.~\citet{Hamilton1988} suggested that brighter galaxies reside in denser environments as a consequence of larger galaxy bias, such that more luminous galaxies form in more dense regions.~\citet{Zehavi2011} and ~\citet{Norberg2002b} show how this bias depends on luminosity and colour.

 Using data from GAMA also allows the LF to be constrained over a range of redshifts, providing a tool with which to measure the evolution of the LF dependence on environment. We find only a very small evolution in the GAMA LF over that already taken into account by the luminosity evolution parameter $Q_0$ (Fig.~\ref{fig:LF_dens_z_data}). This evolution is likely related to the known small amount of density evolution in GAMA~\citep{Loveday2012}. However, the large degeneracies between \Ms~and \phis~make it difficult to determine the variation of \phis~with redshift, and hence we do not try to model any redshift dependent density evolution. We find the value of $Q_0$ to be different for red and blue galaxies. When comparing galaxy properties in different environments it is important to take this into account, since different galaxy populations dominate in different environments (see Fig.~\ref{fig:col_fract}).

Splitting the sample into red and blue galaxies gives an indication of how different populations of galaxies behave in different environments. The left panel of Fig.~\ref{fig:col_fract} shows how the fraction of red and blue galaxies varies with luminosity for different density bins. In general blue galaxies tend to dominate in underdense regions and tend to be fainter, and red galaxies dominate overdense regions and tend to be brighter. This is also seen clearly in Fig.~\ref{fig:schecht_dens_c_data} when considering how \phis~changes with overdensity for red and blue galaxies, and by comparing how the fraction of red galaxies as a function of overdensity (bottom panel) changes with absolute magnitude. Both red and blue samples show a faint-end slope that varies with density for overdense environments only (as Equation~\ref{eqn:alp_dens}), suggesting the suppression of faint galaxies is not as effective in overdense environments when considering a specific galaxy population, but this is not as evident when considering the sample as a whole. The shallower dependence on overdensity seen when considering all galaxies can be attributed to the varying fractions of blue and red populations residing in different environments. This result is in good agreement with the LF found for cluster galaxies in the 2dFGRS ~\citep{DePropris2003}, for which the LF for early type galaxies is found to be considerably steeper in clusters than the LF for field galaxies. A galaxy's local environment has different effects on its colour and morphology (see Figure 8 of~\citealt{Bamford2009}). We expect the morphology-density relation~\citep{Dressler1980} to be similar but not implicitly described by Fig.~\ref{fig:col_fract}.

\subsection{Physical Interpretation}
\label{section:phys_interp}
While the mock catalogues seem to predict a similar overall trend to the data in the shape of the luminosity function for populations of galaxies residing in each environment, there are some significant differences. Fig.~\ref{fig:col_fract} (right panel) shows that the mock catalogues predict that the fraction of red and blue galaxies does not vary as a function of magnitude in the same way as is observed (left panel). Instead, the fraction appears to vary with a much shallower slope for $\Mr>-20.2$, but with a steeper slope for $\Mr < -20.2$. This is true for all environments. The absolute magnitude at which the fraction of blue galaxies and red galaxies are equal gets fainter in denser environments, determining the luminosity at which the dominating population of galaxies changes for a given environment. In the mock catalogues this luminosity is too faint in overdense regions and too bright in the most underdense regions.

A similar discrepancy in the mock catalogues can be seen by comparing the gradient of the fraction of red galaxies as a function of overdensity to GAMA as seen in the bottom panels of Fig.~\ref{fig:schecht_dens_c_data} for different absolute magnitude ranges. For bright galaxies in the approximate range $-20.0<\Mr<-21.0$ the mocks show a red fraction with a shallower dependence on overdensity, such that in the most underdense environments the fraction of galaxies which are red is higher than seen in GAMA. However, for the brightest galaxies the red fraction is predicted to be similar GAMA. For faint galaxies this is the opposite case, the fraction of red galaxies varies with environment more strongly than is seen in GAMA, predicting too many (by up to a factor of two) faint red galaxies in the most overdense environments.

The LF for red galaxies predicted by the mock catalogues is mostly consistent with that measured in GAMA. However, the faint-end slope for red galaxies is predicted to be too steep compared to GAMA by up to $\Delta\alp=0.43$. For blue galaxies the faint-end slope is up to $\Delta\alp=0.58$ shallower in the mock catalogues than in GAMA in overdense regions. The variation of \phis~with environment suggests too many blue galaxies are predicted in overdense environments, slightly too few red galaxies in underdense environments. This discrepancy is reflected in the variation of \Ms~with environment, that is predicted to be weaker than is seen. 

 The shape of the LF for the very bluest galaxies does not seem to show much variation with environment. However, the redder LFs steepen and brighten with overdensity, and this variation is more significant for the intermediate red population (shown by the orange and red curves in the middle panel of Fig.~\ref{fig:LF_dens_c_data}). In general the mock catalogues predict the same result, although it is the reddest population that is seen to vary the most significantly in this case.

The comparison of the LFs of the mock galaxies and GAMA in different environments for different colours is summarised in Fig.~\ref{fig:LFs_gam_mock}. The total LF of GAMA and the mock galaxies when not split by colour or by environment is, by construction, extremely similar. It is therefore not surprising that the LFs in the bottom right panel match particularly well. However, the LFs seem to agree remarkably well when split by environment and colour, barring a few discrepancies. Too many bright galaxies (specifically blue) are predicted in underdense environments. The faint end of the blue LF (which dominates these environments) agrees well, resulting in only a small deviation from the GAMA LF at the faint end in underdense regions. In overdense environments, however, the predicted bright end of the LF is in good agreement with the GAMA LF, and deviations are only apparent in the faint end, where too many faint red galaxies are predicted by the models (as is also visible in Fig.~\ref{fig:schecht_dens_c_data}).

A similar result is found by ~\citet{Baldry2006}, who investigate how the red fraction depends on stellar mass and environment in semi analytical models~\citep{Bower2006,Croton2006} and in SDSS, finding that both models qualitatively agree well with SSDS, particularly the~\citet{Bower2006} model, but that there is an overabundance of red galaxies in more dense regions in both models.

This excess of faint red galaxies in the model can be attributed to the known problem of over-quenching of (dwarf) satellites in most semi-analytical models~\citep{Weinmann2006,Kimm2009}. In the~\citet{Bower2006} model, we find the faint end of the red LF is dominated by satellite galaxies. This is more apparent for the most overdense regions, since the majority of galaxies in overdense regions (massive haloes) are most likely satellite galaxies. Underdense regions are more likely to be occupied by isolated central galaxies, which will evolve with very little environmental influence.

In the~\citet{Bower2006} model, when a galaxy falls into a larger halo and becomes a satellite, its hot gas reservoir is instantaneously lost to the host halo. Once it has depleted its supply of cold gas, star formation will cease. The excess of quenched (red) satellites can be attributed to this too efficient loss of hot gas on infall. Galaxies in isolation (predominantly central galaxies) have their star formation quenched through processes internal to the galaxy and its host halo, for example AGN feedback. By observationally studying how star formation is quenched in different environments, the prescriptions in the models for internal and environmental processes causing quenching can be refined.~\citet{Font2008} incorporated a treatment of stripping of hot gas based on the results of hydrodynamical simulations within the semi-analytic model of~\citet{Bower2006}, to investigate the behaviour of the hot gas reservoir of satellite galaxies. They find that satellite galaxies can retain a significant fraction of their hot gas after infall, allowing them to continue star formation for a significant period of time. This decreases the fraction of red satellite galaxies produced by the model, producing a satellite colour distribution in good agreement with that observed in SDSS.

~\citet{Wheeler2014} find less than 30\% of observed low mass ($M_* \simeq 10^{8.5-9.5}M_{\odot}$) dwarf satellites are quenched, a fraction much lower than is predicted by models, and suggest a long quenching timescale ($>9.5$Gyrs) for satellites of these masses. When comparing these results to those of~\citet{Wetzel2013} and~\citet{DeLucia2012}, who measure a quenching timescale for observed dwarf satellites of higher mass, ~\citet{Wheeler2014} discover the quenching timescale is dependent on stellar mass for satellite galaxies, such that lower stellar mass systems exhibit a longer timescale for quenching star formation. However galaxies also undergo quenching through internal processes, which also correlates strongly with stellar mass. It is likely that these internal processes also contribute to quenching in satellites. When taking this into account,~\citet{Wheeler2014} and~\citet{Wetzel2013} find the fraction of satellites quenched only through environmental processes is independent of stellar mass. 

Taking into account studies of how hot gas is stripped from satellite galaxies on in-fall would help to provide a better model describing the evolution of satellite galaxies. 

Another obvious discrepancy we find between the model and observations is an excess of bright blue galaxies in underdense environments predicted in the model. The majority of galaxies in these environments are centrals, most likely unaffected by processes external to the galaxy (since the number density of galaxies is low). This excess of bright blue galaxies could be due to the halo mass threshold below which AGN feedback is not efficient enough to suppress star formation, allowing for excess blue galaxies to be predicted at the bright end of the LF. The lowest density bin in our sample contains predominantly blue galaxies in halos with masses $M < 10^{12.2}M_{\odot}$.

\section{Conclusion}
\label{section:Conc}
The results presented and discussed above can be summarised as follows.

\begin{itemize}
\item The GAMA galaxy LF varies smoothly with overdensity, such that denser environments contain brighter galaxies, the LF is described by a linear relation between \Ms~and \lonepd. The faint-end slope, \alp, does not show any detectable variation with environment, consistent with results from other galaxy surveys. As expected, $\log_{10}{\phis}$ varies linearly with \lonepd, such that the slope is related to galaxy bias as $1/b_{\rm g}$.
\item When split by colour, the measured LFs confirm that red galaxies dominate overdense environments, and blue galaxies dominate underdense environments. A variation in the faint-end slope with environment becomes apparent, such that \alp~steepens linearly with \lonepd~for $\dddp \ge -0.2$ for red galaxies, but no obvious trend is seen for blue galaxies. The faint-end slope for all galaxies when not split by colour can be understood by considering which colours dominate in which environments.
\item The mock galaxy catalogues constructed from the~\citet{Bower2006} galaxy formation model produce LFs that agree qualitatively with those found in GAMA, when split by environment and by colour. Discrepancies tend to appear in the overabundance of bright blue galaxies predicted by the mock catalogues in underdense environments, which could possibly be attributed to AGN feedback in the lowest mass halos not considered in the model, and the faint end of the red LF in overdense environments, where too many faint red galaxies are predicted. This is likely to be due to hot gas being stripped too efficiently when a galaxy becomes a satellite of in larger halo. 
\end{itemize}

This work will be extended further to investigate results found in this analysis. In particular the availability of various models of galaxy formation, based on those used here, provides a means by which to measure how various aspects of galaxy formation and evolution affect the ability to constrain the galaxy LF in different environments. Comparing the work done here to the work of Eardley et al. in prep, will help to determine whether or not the variation of the LF with environment is due to the local environment in which a galaxy resides, or a more global environment, defined by eg. voids and filaments. The ability to measure galaxy bias through the method described above can also be investigated by measuring how the LF changes with galaxy overdensity for DDP samples covering various magnitude ranges, and for different galaxy populations (eg. colours). The availability of multi-wavelength data as well as stellar masses measured in GAMA, allows for this work to be extended to determine whether or not the trends in the LF seen here are consistent over a larger range of wavelengths or stellar masses.

\section*{Acknowledgments}
TMR acknowledges support from a European Research Council Starting Grant (DEGAS-259586).
PN acknowledges the support of the Royal Society through the award of a University Research Fellowship and the European Research Council, through receipt of a Starting Grant (DEGAS-259586).
This work was supported by the Science and Technology Facilities Council [ST/L00075/1].
Data used in this paper will be available through the GAMA DB ({\tt http://www.gama-survey.org/}) once the associated redshifts are publicly released.
GAMA is a joint European-Australian project based around a spectroscopic campaign using the Anglo-Australian Telescope. The GAMA input catalogue is based on data taken from the Sloan Digital Sky Survey and the UKIRT Infrared Deep Sky Survey. Complementary imaging of the GAMA regions is being obtained by a number of independent survey programs including GALEX MIS, VST KIDS, VISTA VIKING, WISE, Herschel-ATLAS, GMRT and ASKAP providing UV to radio coverage. GAMA is funded by the STFC (UK), the ARC (Australia), the AAO, and the participating institutions. The GAMA website is {\tt http://www.gama-survey.org/}.
This work used the DiRAC Data Centric system at Durham University, operated by the Institute for Computational Cosmology
on behalf of the STFC DiRAC HPC Facility (www.dirac.ac.uk). This equipment was funded by BIS National E-infrastructure
capital grant ST/K00042X/1, STFC capital grant ST/H008519/1, and STFC DiRAC Operations grant ST/K003267/1 and Durham
University. DiRAC is part of the National E-Infrastructure.

\bibliographystyle{mn2e}
\setlength{\bibhang}{2.0em}
\setlength\labelwidth{0.0em}
\bibliography{tamr_lfenvs}

\begin{thebibliography}{62}
\expandafter\ifx\csname natexlab\endcsname\relax\def\natexlab#1{#1}\fi

\bibitem[{{Abazajian} {et~al}\mbox{.}(2009){Abazajian}, {Adelman-McCarthy},
  {Ag{\"u}eros}, {Allam}, {Allende Prieto}, {An}, {Anderson}, {Anderson},
  {Annis}, {Bahcall}, \& et~al.}]{Abazajian2009}
{Abazajian} K.~N. {et~al.}, 2009, \apjs, 182, 543

\bibitem[{{Alpaslan} {et~al}\mbox{.}(2014){Alpaslan}, {Robotham}, {Driver},
  {Norberg}, {Baldry}, {Bauer}, {Bland-Hawthorn}, {Brown}, {Cluver}, {Colless},
  {Foster}, {Hopkins}, {Van Kampen}, {Kelvin}, {Lara-Lopez}, {Liske},
  {Lopez-Sanchez}, {Loveday}, {McNaught-Roberts}, {Merson}, \&
  {Pimbblet}}]{Alpaslan2013}
{Alpaslan} M. {et~al.}, 2014, \mnras, 438, 177

\bibitem[{{Baldry} {et~al}\mbox{.}(2006){Baldry}, {Balogh}, {Bower},
  {Glazebrook}, {Nichol}, {Bamford}, \& {Budavari}}]{Baldry2006}
{Baldry} I.~K., {Balogh} M.~L., {Bower} R.~G., {Glazebrook} K., {Nichol} R.~C.,
  {Bamford} S.~P., {Budavari} T., 2006, \mnras, 373, 469

\bibitem[{{Baldry} {et~al}\mbox{.}(2010){Baldry}, {Robotham}, {Hill}, {Driver},
  {Liske}, {Norberg}, {Bamford}, {Hopkins}, {Loveday}, {Peacock}, {Cameron},
  {Croom}, {Cross}, {Doyle}, {Dye}, {Frenk}, \& {Jones}}]{Baldry2010}
{Baldry} I.~K. {et~al.}, 2010, \mnras, 404, 86

\bibitem[{{Bamford} {et~al}\mbox{.}(2009){Bamford}, {Nichol}, {Baldry}, {Land},
  {Lintott}, {Schawinski}, {Slosar}, {Szalay}, {Thomas}, {Torki}, {Andreescu},
  {Edmondson}, {Miller}, {Murray}, {Raddick}, \& {Vandenberg}}]{Bamford2009}
{Bamford} S.~P. {et~al.}, 2009, \mnras, 393, 1324

\bibitem[{{Benson} {et~al}\mbox{.}(2003{\natexlab{a}}){Benson}, {Bower},
  {Frenk}, {Lacey}, {Baugh}, \& {Cole}}]{Benson2003a}
{Benson} A.~J., {Bower} R.~G., {Frenk} C.~S., {Lacey} C.~G., {Baugh} C.~M.,
  {Cole} S., 2003{\natexlab{a}}, \apj, 599, 38

\bibitem[{{Benson} {et~al}\mbox{.}(2003{\natexlab{b}}){Benson}, {Frenk},
  {Baugh}, {Cole}, \& {Lacey}}]{Benson2003c}
{Benson} A.~J., {Frenk} C.~S., {Baugh} C.~M., {Cole} S., {Lacey} C.~G.,
  2003{\natexlab{b}}, \mnras, 343, 679

\bibitem[{{Benson} {et~al}\mbox{.}(2003{\natexlab{c}}){Benson}, {Hoyle},
  {Torres}, \& {Vogeley}}]{Benson2003b}
{Benson} A.~J., {Hoyle} F., {Torres} F., {Vogeley} M.~S., 2003{\natexlab{c}},
  \mnras, 340, 160

\bibitem[{{Blanton} \& {Berlind}(2007)}]{Blanton2007b}
{Blanton} M.~R., {Berlind} A.~A., 2007, \apj, 664, 791

\bibitem[{{Blanton} {et~al}\mbox{.}(2003{\natexlab{a}}){Blanton}, {Brinkmann},
  {Csabai}, {Doi}, {Eisenstein}, {Fukugita}, {Gunn}, {Hogg}, \&
  {Schlegel}}]{Blanton2003a}
{Blanton} M.~R. {et~al.}, 2003{\natexlab{a}}, \aj, 125, 2348

\bibitem[{{Blanton} {et~al}\mbox{.}(2005){Blanton}, {Eisenstein}, {Hogg},
  {Schlegel}, \& {Brinkmann}}]{Blanton2005}
{Blanton} M.~R., {Eisenstein} D., {Hogg} D.~W., {Schlegel} D.~J., {Brinkmann}
  J., 2005, \apj, 629, 143

\bibitem[{{Blanton} {et~al}\mbox{.}(2003{\natexlab{b}}){Blanton}, {Hogg},
  {Bahcall}, {Brinkmann}, {Britton}, {Connolly}, {Csabai}, {Fukugita},
  {Loveday}, {Meiksin}, {Munn}, {Nichol}, {Okamura}, {Quinn}, {Schneider},
  {Shimasaku}, {Strauss}, {Tegmark}, {Vogeley}, \& {Weinberg}}]{Blanton2003b}
{Blanton} M.~R. {et~al.}, 2003{\natexlab{b}}, \apj, 592, 819

\bibitem[{{Blanton} \& {Roweis}(2007)}]{Blanton2007a}
{Blanton} M.~R., {Roweis} S., 2007, \aj, 133, 734

\bibitem[{{Bower} {et~al}\mbox{.}(2006){Bower}, {Benson}, {Malbon}, {Helly},
  {Frenk}, {Baugh}, {Cole}, \& {Lacey}}]{Bower2006}
{Bower} R.~G., {Benson} A.~J., {Malbon} R., {Helly} J.~C., {Frenk} C.~S.,
  {Baugh} C.~M., {Cole} S., {Lacey} C.~G., 2006, \mnras, 370, 645

\bibitem[{{Bromley} {et~al}\mbox{.}(1998){Bromley}, {Press}, {Lin}, \&
  {Kirshner}}]{Bromley1998}
{Bromley} B.~C., {Press} W.~H., {Lin} H., {Kirshner} R.~P., 1998, \apj, 505, 25

\bibitem[{{Brough} {et~al}\mbox{.}(2013){Brough}, {Croom}, {Sharp}, {Hopkins},
  {Taylor}, {Baldry}, {Gunawardhana}, {Liske}, {Norberg}, {Robotham}, {Bauer},
  {Bland-Hawthorn}, {Colless}, {Foster}, {Kelvin}, {Lara-Lopez},
  {L{\'o}pez-S{\'a}nchez}, {Loveday}, {Owers}, {Pimbblet}, \&
  {Prescott}}]{Brough2013}
{Brough} S. {et~al.}, 2013, \mnras, 435, 2903

\bibitem[{{Colless} {et~al}\mbox{.}(2003){Colless}, {Peterson}, {Jackson},
  {Peacock}, {Cole}, {Norberg}, {Baldry}, {Baugh}, {Bland-Hawthorn}, {Bridges},
  {Cannon}, {Collins}, {Couch}, {Cross}, {Dalton}, {De Propris}, {Driver},
  {Efstathiou}, {Ellis}, {Frenk}, {Glazebrook}, {Lahav}, {Lewis}, {Lumsden},
  {Maddox}, {Madgwick}, {Sutherland}, \& {Taylor}}]{Colless2003}
{Colless} M. {et~al.}, 2003, preprint arXiv:astro-ph/0306581

\bibitem[{{Croton} {et~al}\mbox{.}(2005){Croton}, {Farrar}, {Norberg},
  {Colless}, {Peacock}, {Baldry}, {Baugh}, {Bland-Hawthorn}, {Bridges},
  {Cannon}, {Cole}, {Collins}, {Couch}, {Dalton}, {De Propris}, \&
  {Driver}}]{Croton2005}
{Croton} D.~J. {et~al.}, 2005, \mnras, 356, 1155

\bibitem[{{Croton} {et~al}\mbox{.}(2006){Croton}, {Springel}, {White}, {De
  Lucia}, {Frenk}, {Gao}, {Jenkins}, {Kauffmann}, {Navarro}, \&
  {Yoshida}}]{Croton2006}
{Croton} D.~J. {et~al.}, 2006, \mnras, 365, 11

\bibitem[{{Davis} \& {Huchra}(1982)}]{Davis1982}
{Davis} M., {Huchra} J., 1982, \apj, 254, 437

\bibitem[{{De Lucia} {et~al}\mbox{.}(2012){De Lucia}, {Weinmann}, {Poggianti},
  {Arag{\'o}n-Salamanca}, \& {Zaritsky}}]{DeLucia2012}
{De Lucia} G., {Weinmann} S., {Poggianti} B.~M., {Arag{\'o}n-Salamanca} A.,
  {Zaritsky} D., 2012, \mnras, 423, 1277

\bibitem[{{De Propris} {et~al}\mbox{.}(2003){De Propris}, {Colless}, {Driver},
  {Couch}, {Peacock}, {Baldry}, {Baugh}, {Bland-Hawthorn}, {Bridges}, {Cannon},
  {Cole}, {Collins}, {Cross}, {Dalton}, {Efstathiou}, {Ellis}, {Frenk},
  {Glazebrook}, {Hawkins}, {Jackson}, {Lahav}, {Lewis}, {Lumsden}, {Maddox},
  {Madgwick}, {Norberg}, {Percival}, {Peterson}, {Sutherland}, \&
  {Taylor}}]{DePropris2003}
{De Propris} R. {et~al.}, 2003, \mnras, 342, 725

\bibitem[{{De Propris} {et~al}\mbox{.}(2013){De Propris}, {Phillipps}, \&
  {Bremer}}]{DePropris2013}
{De Propris} R., {Phillipps} S., {Bremer} M.~N., 2013, \mnras, 434, 3469

\bibitem[{{Dressler}(1980)}]{Dressler1980}
{Dressler} A., 1980, \apj, 236, 351

\bibitem[{{Driver} {et~al}\mbox{.}(2011){Driver}, {Hill}, {Kelvin}, {Robotham},
  {Liske}, {Norberg}, {Baldry}, {Bamford}, {Hopkins}, {Loveday}, {Peacock},
  {Andrae}, {Bland-Hawthorn}, {Brough}, {Brown}, {Cameron}, {Ching}, \&
  {Colless}}]{Driver2011}
{Driver} S.~P. {et~al.}, 2011, \mnras, 413, 971

\bibitem[{{Einasto} {et~al}\mbox{.}(2005){Einasto}, {Suhhonenko},
  {Hein{\"a}m{\"a}ki}, {Einasto}, \& {Saar}}]{Einasto2005}
{Einasto} M., {Suhhonenko} I., {Hein{\"a}m{\"a}ki} P., {Einasto} J., {Saar} E.,
  2005, \aap, 436, 17

\bibitem[{{Eke} {et~al}\mbox{.}(2004){Eke}, {Frenk}, {Baugh}, {Cole},
  {Norberg}, {Peacock}, {Baldry}, {Bland-Hawthorn}, {Bridges}, {Cannon},
  {Colless}, {Collins}, {Couch}, {Dalton}, {de Propris}, {Driver},
  {Efstathiou}, {Ellis}, {Glazebrook}, {Jackson}, {Lahav}, {Lewis}, {Lumsden},
  {Maddox}, {Madgwick}, {Peterson}, {Sutherland}, \& {Taylor}}]{Eke2004}
{Eke} V.~R. {et~al.}, 2004, \mnras, 355, 769

\bibitem[{{Font} {et~al}\mbox{.}(2008){Font}, {Bower}, {McCarthy}, {Benson},
  {Frenk}, {Helly}, {Lacey}, {Baugh}, \& {Cole}}]{Font2008}
{Font} A.~S. {et~al.}, 2008, \mnras, 389, 1619

\bibitem[{{Gunawardhana} {et~al}\mbox{.}(2013){Gunawardhana}, {Hopkins},
  {Bland-Hawthorn}, {Brough}, {Sharp}, {Loveday}, {Taylor}, {Jones},
  {Lara-L{\'o}pez}, {Bauer}, {Colless}, {Owers}, {Baldry},
  {L{\'o}pez-S{\'a}nchez}, {Foster}, {Bamford}, {Brown}, {Driver},
  {Drinkwater}, {Liske}, {Meyer}, {Norberg}, {Robotham}, {Ching}, {Cluver},
  {Croom}, {Kelvin}, {Prescott}, {Steele}, {Thomas}, \&
  {Wang}}]{Gunawardhana2013}
{Gunawardhana} M.~L.~P. {et~al.}, 2013, \mnras, 433, 2764

\bibitem[{{Hamilton}(1988)}]{Hamilton1988}
{Hamilton} A.~J.~S., 1988, \apjl, 331, L59

\bibitem[{{Hopkins} {et~al}\mbox{.}(2013){Hopkins}, {Driver}, {Brough},
  {Owers}, {Bauer}, {Gunawardhana}, {Cluver}, {Colless}, {Foster},
  {Lara-L{\'o}pez}, {Roseboom}, {Sharp}, {Steele}, {Thomas}, {Baldry}, {Brown},
  {Liske}, {Norberg}, {Robotham}, {Bamford}, {Bland-Hawthorn}, {Drinkwater},
  {Loveday}, {Meyer}, {Peacock}, {Tuffs}, {Agius}, {Alpaslan}, {Andrae},
  {Cameron}, {Cole}, {Ching}, {Christodoulou}, {Conselice}, {Croom}, {Cross},
  {De Propris}, {Delhaize}, {Dunne}, {Eales}, {Ellis}, {Frenk}, {Graham},
  {Grootes}, {H{\"a}u{\ss}ler}, {Heymans}, {Hill}, {Hoyle}, {Hudson}, {Jarvis},
  {Johansson}, {Jones}, {van Kampen}, {Kelvin}, {Kuijken},
  {L{\'o}pez-S{\'a}nchez}, {Maddox}, {Madore}, {Maraston}, {McNaught-Roberts},
  {Nichol}, {Oliver}, {Parkinson}, {Penny}, {Phillipps}, {Pimbblet}, {Ponman},
  {Popescu}, {Prescott}, {Proctor}, {Sadler}, {Sansom}, {Seibert},
  {Staveley-Smith}, {Sutherland}, {Taylor}, {Van Waerbeke}, {V{\'a}zquez-Mata},
  {Warren}, {Wijesinghe}, {Wild}, \& {Wilkins}}]{Hopkins2013}
{Hopkins} A.~M. {et~al.}, 2013, \mnras, 430, 2047

\bibitem[{{Hoyle} {et~al}\mbox{.}(2005){Hoyle}, {Rojas}, {Vogeley}, \&
  {Brinkmann}}]{Hoyle2005}
{Hoyle} F., {Rojas} R.~R., {Vogeley} M.~S., {Brinkmann} J., 2005, \apj, 620,
  618

\bibitem[{{H{\"u}tsi} {et~al}\mbox{.}(2002){H{\"u}tsi}, {Einasto}, {Tucker},
  {Saar}, {Einasto}, {M{\"u}ller}, {Hein{\"a}m{\"a}ki}, \& {Allam}}]{Hutsi2002}
{H{\"u}tsi} G., {Einasto} J., {Tucker} D.~L., {Saar} E., {Einasto} M.,
  {M{\"u}ller} V., {Hein{\"a}m{\"a}ki} P., {Allam} S.~S., 2002, preprint
  arXiv:astro-ph/0212327

\bibitem[{{Kimm} {et~al}\mbox{.}(2009){Kimm}, {Somerville}, {Yi}, {van den
  Bosch}, {Salim}, {Fontanot}, {Monaco}, {Mo}, {Pasquali}, {Rich}, \&
  {Yang}}]{Kimm2009}
{Kimm} T. {et~al.}, 2009, \mnras, 394, 1131

\bibitem[{{Lin} {et~al}\mbox{.}(1996){Lin}, {Kirshner}, {Shectman}, {Landy},
  {Oemler}, {Tucker}, \& {Schechter}}]{Lin1996}
{Lin} H., {Kirshner} R.~P., {Shectman} S.~A., {Landy} S.~D., {Oemler} A.,
  {Tucker} D.~L., {Schechter} P.~L., 1996, \apj, 464, 60

\bibitem[{{Loveday} {et~al}\mbox{.}(2012){Loveday}, {Norberg}, {Baldry},
  {Driver}, {Hopkins}, {Peacock}, {Bamford}, {Liske}, {Bland-Hawthorn},
  {Brough}, {Brown}, {Cameron}, {Conselice}, {Croom}, {Frenk}, {Gunawardhana},
  \& {Hill}}]{Loveday2012}
{Loveday} J. {et~al.}, 2012, \mnras, 420, 1239

\bibitem[{{Madgwick} {et~al}\mbox{.}(2002){Madgwick}, {Lahav}, {Baldry},
  {Baugh}, {Bland-Hawthorn}, {Bridges}, {Cannon}, {Cole}, {Colless}, {Collins},
  {Couch}, {Dalton}, {De Propris}, {Driver}, {Efstathiou}, {Ellis}, {Frenk},
  {Glazebrook}, {Jackson}, {Lewis}, {Lumsden}, {Maddox}, {Norberg}, {Peacock},
  {Peterson}, {Sutherland}, \& {Taylor}}]{Madgwick2002}
{Madgwick} D.~S. {et~al.}, 2002, \mnras, 333, 133

\bibitem[{{Mahajan} \& {Raychaudhury}(2009)}]{Mahajan2009}
{Mahajan} S., {Raychaudhury} S., 2009, \mnras, 400, 687

\bibitem[{{Maller} {et~al}\mbox{.}(2009){Maller}, {Berlind}, {Blanton}, \&
  {Hogg}}]{Maller2009}
{Maller} A.~H., {Berlind} A.~A., {Blanton} M.~R., {Hogg} D.~W., 2009, \apj,
  691, 394

\bibitem[{{Mathis} \& {White}(2002)}]{MathisWhite2002}
{Mathis} H., {White} S.~D.~M., 2002, \mnras, 337, 1193

\bibitem[{{Merson} {et~al}\mbox{.}(2013){Merson}, {Baugh}, {Helly},
  {Gonzalez-Perez}, {Cole}, {Bielby}, {Norberg}, {Frenk}, {Benson}, {Bower},
  {Lacey}, \& {Lagos}}]{Merson2013}
{Merson} A.~I. {et~al.}, 2013, \mnras, 429, 556

\bibitem[{{Mo} {et~al}\mbox{.}(2004){Mo}, {Yang}, {van den Bosch}, \&
  {Jing}}]{Mo2004}
{Mo} H.~J., {Yang} X., {van den Bosch} F.~C., {Jing} Y.~P., 2004, \mnras, 349,
  205

\bibitem[{{Muldrew} {et~al}\mbox{.}(2012){Muldrew}, {Croton}, {Skibba},
  {Pearce}, {Ann}, {Baldry}, {Brough}, {Choi}, {Conselice}, {Cowan},
  {Gallazzi}, {Gray}, {Gr{\"u}tzbauch}, {Li}, {Park}, {Pilipenko}, {Podgorzec},
  \& {Robotham}}]{Muldrew2012}
{Muldrew} S.~I. {et~al.}, 2012, \mnras, 419, 2670

\bibitem[{{Norberg} {et~al}\mbox{.}(2002{\natexlab{a}}){Norberg}, {Baugh},
  {Hawkins}, {Maddox}, {Madgwick}, {Lahav}, {Cole}, {Frenk}, {Baldry},
  {Bland-Hawthorn}, {Bridges}, {Cannon}, {Colless}, {Collins}, {Couch},
  {Dalton}, {De Propris}, {Driver}, {Efstathiou}, {Ellis}, {Glazebrook},
  {Jackson}, {Lewis}, {Lumsden}, {Peacock}, {Peterson}, {Sutherland}, \&
  {Taylor}}]{Norberg2002b}
{Norberg} P. {et~al.}, 2002{\natexlab{a}}, \mnras, 332, 827

\bibitem[{{Norberg} {et~al}\mbox{.}(2002{\natexlab{b}}){Norberg}, {Cole},
  {Baugh}, {Frenk}, {Baldry}, {Bland-Hawthorn}, {Bridges}, {Cannon}, {Colless},
  {Collins}, {Couch}, {Cross}, {Dalton}, {De Propris}, {Driver}, {Efstathiou},
  {Ellis}, {Glazebrook}, {Jackson}, {Lahav}, {Lewis}, {Lumsden}, {Maddox},
  {Madgwick}, {Peacock}, {Peterson}, {Sutherland}, {Taylor}, \& {2DFGRS
  Team}}]{Norberg2002a}
{Norberg} P. {et~al.}, 2002{\natexlab{b}}, \mnras, 336, 907

\bibitem[{{Peebles}(2001)}]{Peebles2001}
{Peebles} P.~J.~E., 2001, \apj, 557, 495

\bibitem[{{Robotham} {et~al}\mbox{.}(2010{\natexlab{a}}){Robotham}, {Driver},
  {Norberg}, {Baldry}, {Bamford}, {Hopkins}, {Liske}, {Loveday}, {Peacock},
  {Cameron}, {Croom}, {Doyle}, {Frenk}, {Hill}, {Jones}, {van Kampen},
  {Kelvin}, \& {Kuijken}}]{Robotham2010a}
{Robotham} A. {et~al.}, 2010{\natexlab{a}}, \pasa, 27, 76

\bibitem[{{Robotham} {et~al}\mbox{.}(2010{\natexlab{b}}){Robotham},
  {Phillipps}, \& {de Propris}}]{Robotham2010b}
{Robotham} A., {Phillipps} S., {de Propris} R., 2010{\natexlab{b}}, \mnras,
  403, 1812

\bibitem[{{Robotham} {et~al}\mbox{.}(2006){Robotham}, {Wallace}, {Phillipps},
  \& {De Propris}}]{Robotham2006}
{Robotham} A., {Wallace} C., {Phillipps} S., {De Propris} R., 2006, \apj, 652,
  1077

\bibitem[{{Robotham} {et~al}\mbox{.}(2011){Robotham}, {Norberg}, {Driver},
  {Baldry}, {Bamford}, {Hopkins}, {Liske}, {Loveday}, {Merson}, {Peacock},
  {Brough}, {Cameron}, {Conselice}, {Croom}, {Frenk}, {Gunawardhana}, {Hill},
  {Jones}, {Kelvin}, {Kuijken}, {Nichol}, {Parkinson}, {Pimbblet}, {Phillipps},
  {Popescu}, {Prescott}, {Sharp}, {Sutherland}, {Taylor}, {Thomas}, {Tuffs},
  {van Kampen}, \& {Wijesinghe}}]{Robotham2011}
{Robotham} A.~S.~G. {et~al.}, 2011, \mnras, 416, 2640

\bibitem[{{Schechter}(1976)}]{Schechter1976}
{Schechter} P., 1976, \apj, 203, 297

\bibitem[{{Springel} {et~al}\mbox{.}(2005){Springel}, {White}, {Jenkins},
  {Frenk}, {Yoshida}, {Gao}, {Navarro}, {Thacker}, {Croton}, {Helly},
  {Peacock}, {Cole}, {Thomas}, {Couchman}, {Evrard}, {Colberg}, \&
  {Pearce}}]{Springel2005}
{Springel} V. {et~al.}, 2005, \nat, 435, 629

\bibitem[{{Tempel} {et~al}\mbox{.}(2011){Tempel}, {Saar}, {Liivam{\"a}gi},
  {Tamm}, {Einasto}, {Einasto}, \& {M{\"u}ller}}]{Tempel2011}
{Tempel} E., {Saar} E., {Liivam{\"a}gi} L.~J., {Tamm} A., {Einasto} J.,
  {Einasto} M., {M{\"u}ller} V., 2011, \aap, 529, A53

\bibitem[{{Verdes-Montenegro} {et~al}\mbox{.}(2005){Verdes-Montenegro},
  {Sulentic}, {Lisenfeld}, {Leon}, {Espada}, {Garcia}, {Sabater}, \&
  {Verley}}]{Verdes-Montenegro2005}
{Verdes-Montenegro} L., {Sulentic} J., {Lisenfeld} U., {Leon} S., {Espada} D.,
  {Garcia} E., {Sabater} J., {Verley} S., 2005, \aap, 436, 443

\bibitem[{{Weinmann} {et~al}\mbox{.}(2006){Weinmann}, {van den Bosch}, {Yang},
  {Mo}, {Croton}, \& {Moore}}]{Weinmann2006}
{Weinmann} S.~M., {van den Bosch} F.~C., {Yang} X., {Mo} H.~J., {Croton} D.~J.,
  {Moore} B., 2006, \mnras, 372, 1161

\bibitem[{{Wetzel} {et~al}\mbox{.}(2012){Wetzel}, {Tinker}, \&
  {Conroy}}]{Wetzel2012}
{Wetzel} A.~R., {Tinker} J.~L., {Conroy} C., 2012, \mnras, 424, 232

\bibitem[{{Wetzel} {et~al}\mbox{.}(2013){Wetzel}, {Tinker}, {Conroy}, \& {van
  den Bosch}}]{Wetzel2013}
{Wetzel} A.~R., {Tinker} J.~L., {Conroy} C., {van den Bosch} F.~C., 2013,
  \mnras, 432, 336

\bibitem[{{Wheeler} {et~al}\mbox{.}(2014){Wheeler}, {Phillips}, {Cooper},
  {Boylan-Kolchin}, \& {Bullock}}]{Wheeler2014}
{Wheeler} C., {Phillips} J.~I., {Cooper} M.~C., {Boylan-Kolchin} M., {Bullock}
  J.~S., 2014, \mnras, 442, 1396

\bibitem[{{Wijesinghe} {et~al}\mbox{.}(2012){Wijesinghe}, {Hopkins}, {Brough},
  {Taylor}, {Norberg}, {Bauer}, {Brown}, {Cameron}, {Conselice}, {Croom},
  {Driver}, {Grootes}, {Jones}, {Kelvin}, {Loveday}, {Pimbblet}, {Popescu},
  {Prescott}, {Sharp}, {Baldry}, {Sadler}, {Liske}, {Robotham}, {Bamford},
  {Bland-Hawthorn}, {Gunawardhana}, {Meyer}, {Parkinson}, {Drinkwater},
  {Peacock}, \& {Tuffs}}]{Wijesinghe2012}
{Wijesinghe} D.~B. {et~al.}, 2012, \mnras, 423, 3679

\bibitem[{{Yang} {et~al}\mbox{.}(2005){Yang}, {Mo}, {van den Bosch}, \&
  {Jing}}]{Yang2005}
{Yang} X., {Mo} H.~J., {van den Bosch} F.~C., {Jing} Y.~P., 2005, \mnras, 356,
  1293

\bibitem[{{Zandivarez} {et~al}\mbox{.}(2006){Zandivarez}, {Mart{\'{\i}}nez}, \&
  {Merch{\'a}n}}]{Zandivarez2006}
{Zandivarez} A., {Mart{\'{\i}}nez} H.~J., {Merch{\'a}n} M.~E., 2006, \apj, 650,
  137

\bibitem[{{Zehavi} {et~al}\mbox{.}(2011){Zehavi}, {Zheng}, {Weinberg},
  {Blanton}, {Bahcall}, {Berlind}, {Brinkmann}, {Frieman}, {Gunn}, {Lupton},
  {Nichol}, {Percival}, {Schneider}, {Skibba}, {Strauss}, {Tegmark}, \&
  {York}}]{Zehavi2011}
{Zehavi} I. {et~al.}, 2011, \apj, 736, 59

\end{thebibliography}

\begin{appendix}

\section{Luminosity evolution correction, $Q_0$}
\label{appendix:Q_0}
To quantify luminosity evolution in the galaxy population, the GAMA-II data-set is split into 3 redshift bins: $0.01<z<0.21$, $0.21<z<0.31$, $0.31<z<0.51$. The luminosity function is measured for each of these ranges, originally assuming no luminosity evolution ($Q_0 = 0$). When fitting a Schechter function to the LFs at higher redshifts, the faint-end slope, \alp, is not well constrained. Similarly we cannot reliably measure evolution in \phis~using this method. Therefore, for the higher redshifts, \alp~and \phis~are fixed to the values found for the lowest redshift bin. Jackknife errors are used to determine uncertainties on the LF. The value of $Q_0$ can then be estimated by measuring the increase in \Ms~with redshift. Again the uncertainty on \Ms~is found using jackknife errors. The new value for $Q_0$ is used to again measure the LF in the 3 redshift bins, and repeat the process iterating on $Q_0$ until the difference between subsequent values of $Q_0$ is less than 0.01.

This process is carried out for red and blue galaxies in order to determine luminosity evolution for the different populations. $Q_{0,\rm red}$ and $Q_{0,\rm blue}$ are used when measuring LFs.

The values found for $Q_{0,\rm red}$ and $Q_{0,\rm blue}$ (given in \S\ref{subsection:GAMA}) are significantly different from those found in~\citet{Loveday2012}, mostly due to our assumption of no density evolution, $P_0=0$. Density and luminosity evolution are highly degenerate~\citep{Loveday2012}, and therefore not allowing \phis~to vary with redshift allows much different values for $Q_0$. However, the redshift range used in this analysis is not large enough to allow for a small change in $Q_0$ to significantly affect the shape of the LF.

\section{DDP comparison}
\label{appendix:DDP}

\begin{figure}
\centering
\includegraphics[width=85mm]{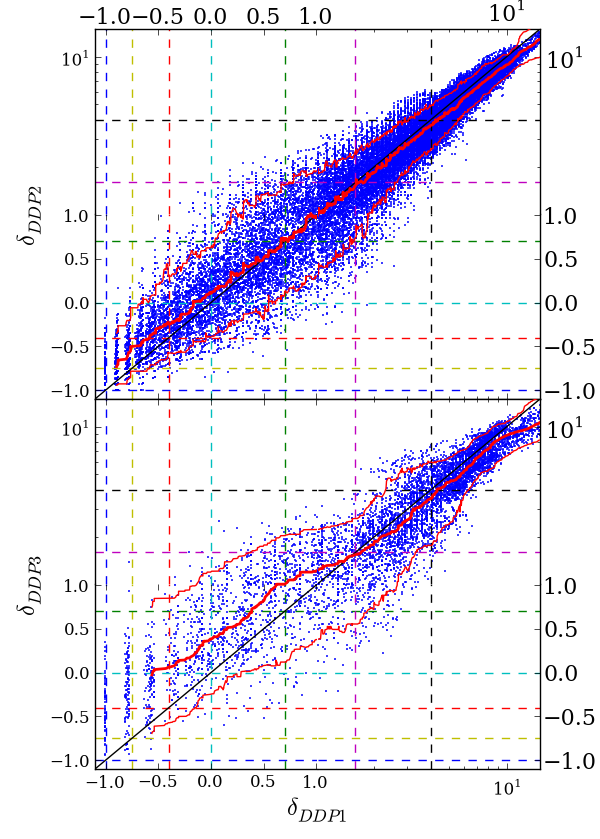}
\caption{\small{Comparison of overdensities measured by different DDP samples. Top panel compares DDP2 overdensities to DDP1 overdensities, for galaxies in the common redshift range to both DDP samples. The running median, 10th and 90th percentiles are shown by the solid and dashed thick, red lines. The lower panel shows a similar comparison, but for DDP3 and DDP1. The chosen overdensity bin limits are shown by the coloured dashed lines (using the same colour coding as in Fig.~\ref{fig:M_delta}).}}
\label{fig:DDP_comp}
\end{figure}

\begin{figure*}
\centering
\includegraphics[width=150mm]{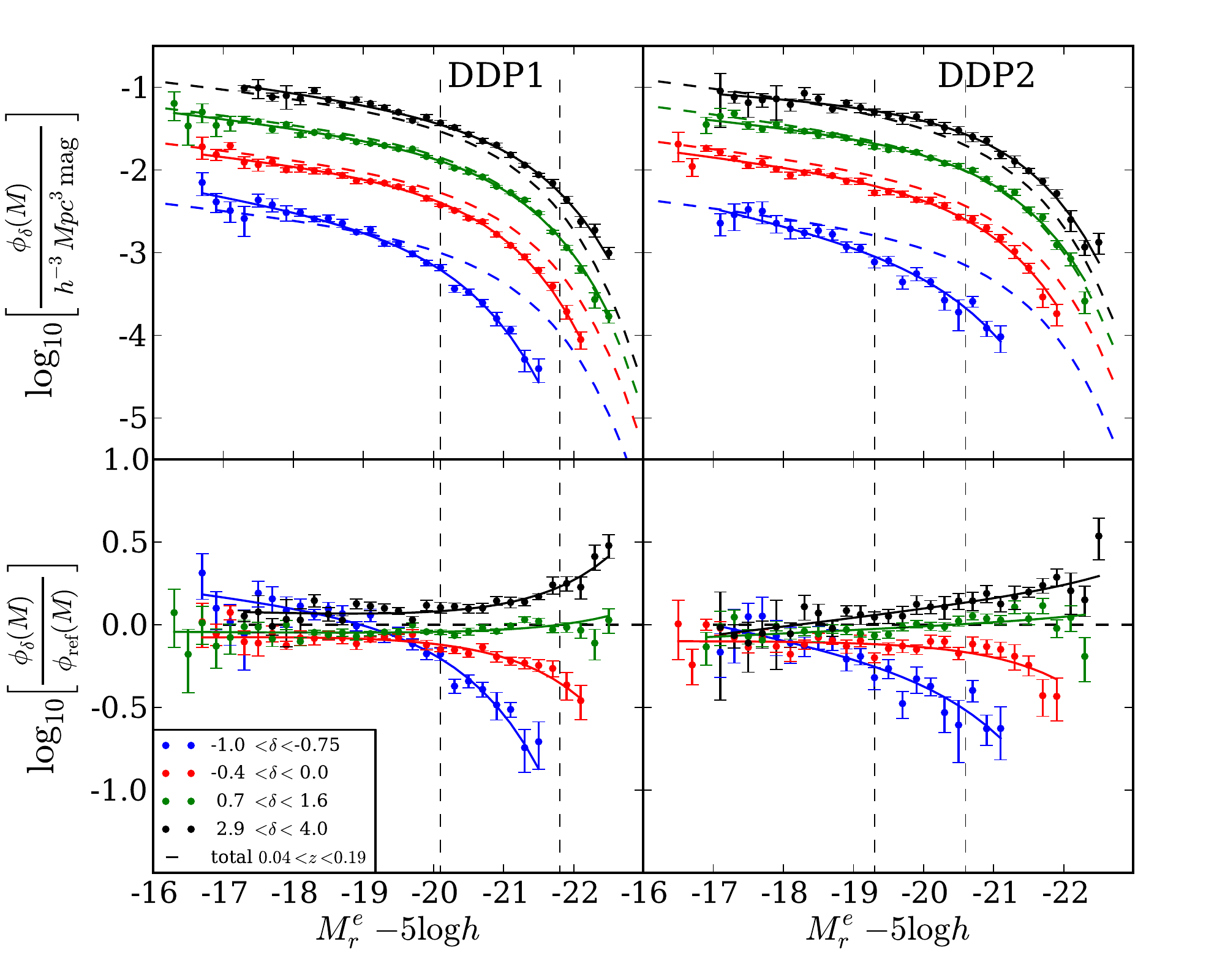}
\caption{\small{LFs for different overdensity bins in GAMA, with Schechter function fits, for overdensity measures using DDP1 (left panel) and DDP2 (right panel), within the redshift range covered by both DDP1 and DDP2 ($0.04<z<0.19$). Bottom panels show the ratio to the total Schechter function fit for the redshift range (as in Fig.~\ref{fig:LF_dens_DDPs_data}). The different tracers of environment do not show significant differences in the resulting LF, although the LFs in extremely underdense bins tend to be underestimated when using DDP2 in comparison to DDP1. Assuming the absolute magnitude range given by DDP2 gives a more reliable representation of the underlying density distribution, the small variation shown here suggests that DDP1 is an acceptable alternative to DDP2.}}
\label{fig:LF_dens_DDP12_data}
\end{figure*}

The precise definition used for the density classification could potentially have a quantitative effect on the results obtained. In this appendix we address whether or not there is a qualitative effect that needs to be accounted for.

Brighter galaxies tend to live in more overdense regions (and higher mass halos, e.g.~\citealt{Einasto2005}), whereas underdense regions (lower mass halos) are populated with fainter galaxies (e.g.~\citealt{Hamilton1988,Zandivarez2006}). Due to this strong correlation between absolute magnitude and environment, it is possible that a DDP sample containing bright galaxies would be biased towards overdense environments~\citep{Zehavi2011}, thereby sampling a particularly large dynamic range of overdense environments compared to an unbiased sample of galaxy tracers and a smaller range in underdense environments.

Fig.~\ref{fig:DDP_comp} shows how the overdensity depends on the DDP sample used. The top panel shows galaxies in the redshift range covered by both DDP1 and DDP2 ($0.04<z<0.19$), with overdensities measured by DDP1 and DDP2 on the x-axis and y-axis respectively. Both DDPs measure extremely similar overdensities, shown by the median of the galaxies as a function of DDP2 (thick red line), with the 10th and 90th percentiles (dashed red line) showing the scatter does not typically extend to more than an overdensity bin (where overdensity bins are shown by coloured dashed lines). The lower panel compares $\delta_{\rm DDP3}$ with $\delta_{\rm DDP1}$ over their common redshift range ($0.04<z<0.10$). The median shows the overdensities measured are very similar. However, below $\dddp=1$ (lower left of the figure), DDP3 tracers seem to measure higher overdensities than DDP1, and above $\dddp=1$ (upper right), DDP3 traces slightly underestimate overdensities in comparison to DDP1. 

Therefore when measuring overdensities for galaxies, it is important to note that the sample used to trace density can have an impact on which galaxies fall into the most underdense density bins.

Fig.~\ref{fig:LF_dens_DDP12_data} shows how the LF changes for overdensities measured by DDP1 (left) and DDP2 (right). The shape of the LF does not vary significantly depending on which DDP sample is used to measure overdensity, suggesting DDP tracers allow for a robust measure of overdensity.

\section{Completeness threshold}

\begin{figure}
\centering
\includegraphics[width=88mm]{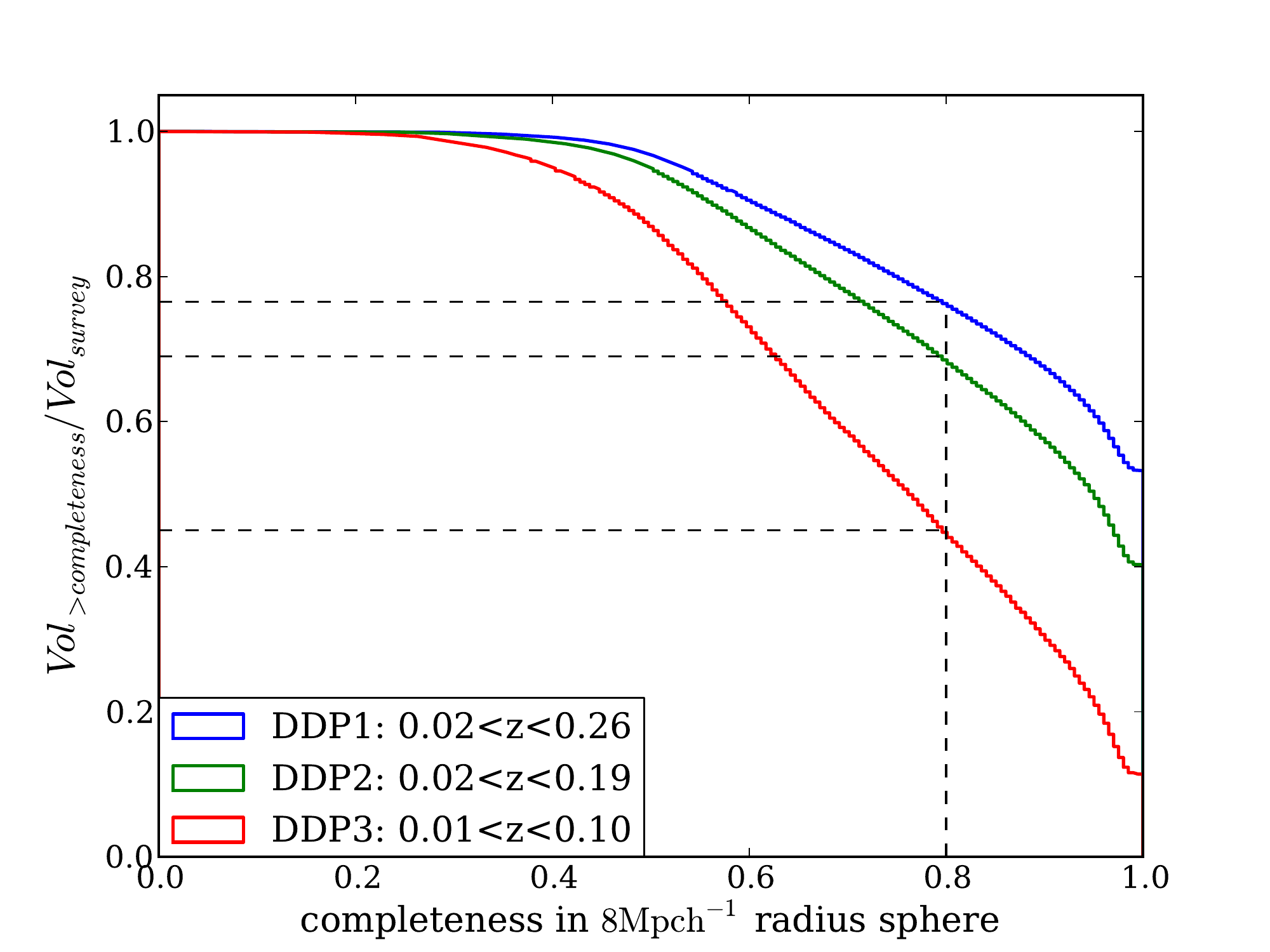}
\caption{\small{Fraction of volume retained in the sample as a function of spectroscopic and masking completeness threshold. A completeness threshold of 80\% retains 77\% of DDP1, but only 45\% of DDP3. If a 4\Mpc radius sphere was used rather than 8\Mpcns, 89\% of DDP1 would be retained for the same completeness threshold.}}
\label{fig:comp_z}
\end{figure}

\label{appendix:comp}
To ensure robust results, a completeness threshold is set to discard galaxies for which the completeness correction is large. Fig.~\ref{fig:comp_z} shows how the fraction of the volume of galaxies kept in the sample decreases as a function of the completeness threshold chosen, for the 3 different DDP samples shown in Fig.~\ref{fig:M_z}. The denser (and hence fainter) the DDP sample is, the smaller the redshift range is and hence the larger the volume correction becomes with the completeness threshold applied. A completeness threshold of 80\% (as adopted here) retains 77\% of the volume of the sample defined by DDP1.

\section{Degeneracies in \Ms~and \alp}
\begin{figure}
\centering
\includegraphics[width=88mm]{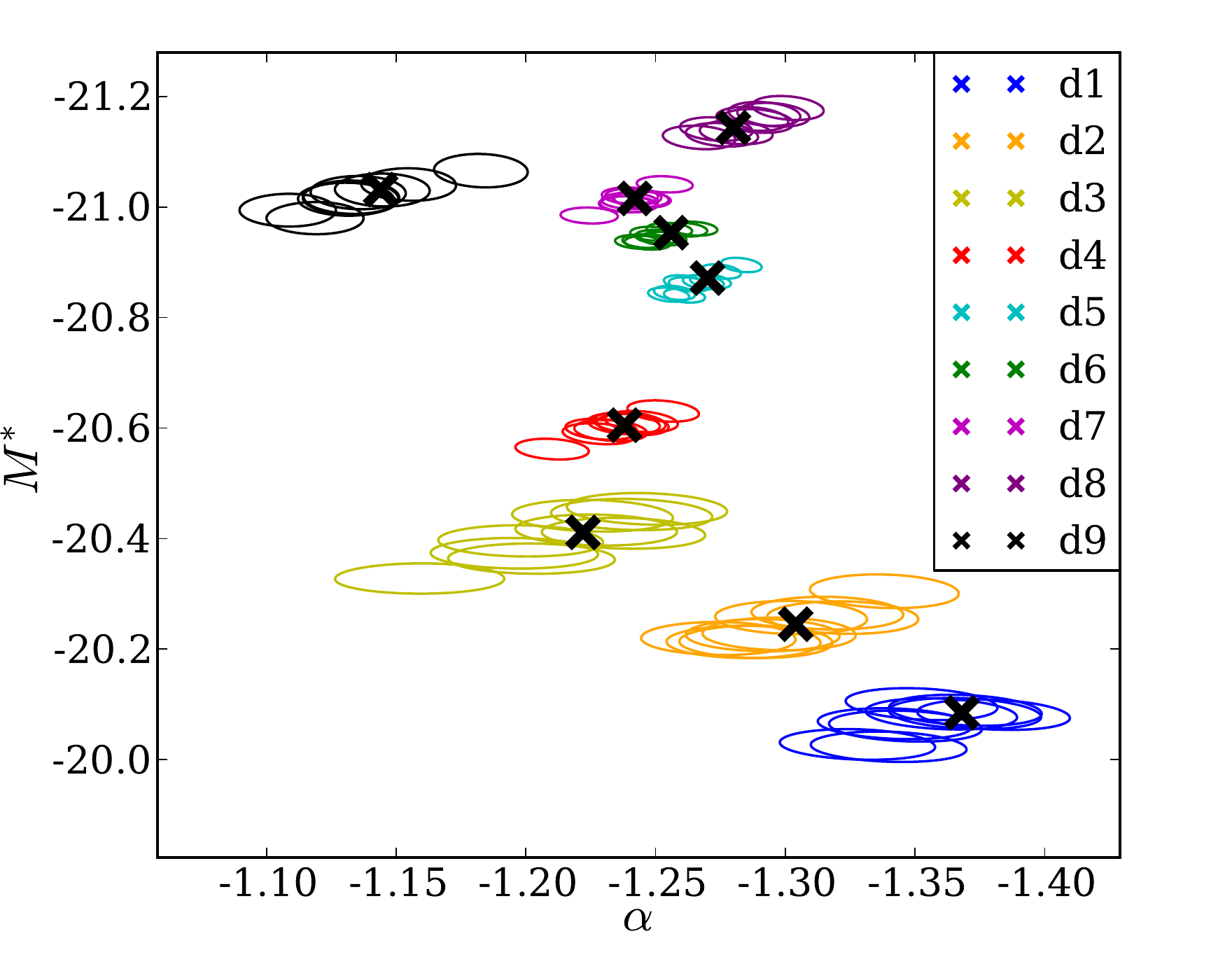}
\caption{\small{1$\sigma$ contours in the \Ms-\alp~plane for each jackknife sample for all 9 density bins in GAMA, coloured by density bin. The best fit value for the total sample is shown by the black crosses in each density bin. The degeneracies between \alp~and \Ms~are obvious within a given density bin.}}
\label{fig:Ms_alp_cont}
\end{figure}

\label{appendix:Ms_alp_deg}
There are well known degeneracies in the parameters that define the Schechter function, \alp, \Ms~and \phis. These degeneracies make it difficult to determine whether or not an apparent trend in any of these parameters with overdensity is true. Fig.~\ref{fig:Ms_alp_cont} shows $1\sigma$ contours for the 9 jackknife samples within each density bin. A brightening of \Ms~by 0.1 mag corresponds to a steepening of \alp~by $\sim0.07$. The offset of the contours confirms our result that the parameters vary strongly with environment. This clear variation of the \Ms~- \alp~degeneracy with environment is also shown in Fig. 6 of~\citet{Croton2005}.

\end{appendix}

\label{lastpage}
\clearpage
\end{document}